\documentclass[a4paper,onecolumn,unpublished,noarxiv]{quantumarticle}
\pdfoutput=1

\usepackage{graphicx}
\usepackage{amssymb}
\usepackage{mathtools}
\usepackage{amsmath}
\usepackage{hyperref}
\usepackage{dsfont}
\usepackage{amsthm}
\usepackage{empheq}
\usepackage{caption}
\usepackage{tabularx}
\usepackage{enumitem}
\usepackage{tikz}
\usetikzlibrary{calc,arrows.meta,patterns,decorations.pathreplacing,decorations.pathmorphing}
\usepackage{quantikz}
\usepackage{subcaption}
\usepackage{array}
\usepackage{booktabs}
\usepackage{float}

\usepackage[sorting=none]{biblatex}
\usepackage{xurl}
\usepackage{microtype}
\providecommand{\ket}[1]{|#1\rangle}
\providecommand{\bra}[1]{\langle#1|}
\providecommand{\proj}[1]{|#1\rangle\!\langle#1|}

\providecommand{\tr}{\operatorname{tr}}

\providecommand{\I}{\mathds{1}}

\providecommand{\nn}{\nonumber}

\newcolumntype{L}[1]{>{\raggedright\arraybackslash}p{#1}}

\theoremstyle{definition}

\newtheorem{theorem}{Theorem}

\newtheorem{proposition}[theorem]{Proposition}

\title{Scaling Laws of Quantum Networks: An Entanglement Transport Framework}

\author{Jaemin Kim}
\affiliation{Department of Electronic Systems, Aalborg University, Fredrik Bajers Vej 7C, 9220 Aalborg, Denmark}
\orcid{0009-0007-1877-3107}
\email{jaemink@es.aau.dk}

\author{Petar Popovski}
\affiliation{Department of Electronic Systems, Aalborg University, Fredrik Bajers Vej 7C, 9220 Aalborg, Denmark}
\orcid{0000-0001-6195-4797}
\email{petarp@es.aau.dk}

\date{\today}

\begin{document}

\maketitle

\begin{abstract}
As quantum networks move toward multi-user architectures, a central question is how the performance of entanglement distribution scales with the numbers of network nodes \(N\) and active users \(M\), yet existing studies use different objectives, resource assumptions, and normalizations.
To account for delivery distance under multi-user demand, we define \emph{entanglement transport}, which weights each delivered Bell pair meeting a fixed fidelity threshold by its endpoint separation.
In a two-layer framework, (i) a \emph{resource layer} measures capacity per activation slot under different aggregation regimes, and (ii) a \emph{physical layer} measures service per second under implementation constraints.
For matched implementations with a bounded number of activation slots per second, asymptotic-aggregation capacity sets a service ceiling, whereas achievable service must be established separately.
On expanding honeycomb lattices with one random session, fixed-window entanglement-percolation constructions achieve \(\Theta(N^{-1})\) entanglement transport per activation slot below their thresholds and \(\Theta(\sqrt N)\) above them.
For homogeneous chains, asymptotic aggregation exceeds fixed-window capacity by a factor \(\Theta(N^2/M)\), and in spatial expansion a noisy physical implementation achieves \(\Theta(M/N)\) service per second.
These examples show that entanglement-transport scaling depends not only on connectivity but also on resource aggregation and physical implementation.
The framework provides a common basis on which quantum-network scaling laws can be established.
\end{abstract}

\section{Introduction}
\label{sec:introduction}

Quantum networks are envisioned as a shared communication infrastructure connecting many geographically separated users, extending quantum communication beyond dedicated point-to-point links~\cite{kimble2008quantum,wehner2018quantum}.
By distributing entanglement across such networks, they can support tasks including secure communication, distributed quantum computation, and distributed sensing~\cite{wehner2018quantum,lee2024quantum}.
Experiments have progressed from two-node links to a three-node network with entanglement swapping through an intermediate node~\cite{pompili2021realization} and to memory--memory entanglement over deployed metropolitan fiber~\cite{knaut2024entanglement}.
As research progresses from elementary links and repeater chains toward routing, network capacities, utility metrics, and hardware-aware protocols, a central question is how network-level performance scales as the infrastructure and the population of active users grow.

The entanglement-generation rate between a fixed pair of users is an elementary benchmark, but it does not by itself characterize a shared network serving many user pairs over different physical distances.
Under multi-user demand, the number of delivered pairs can be increased by serving nearby user pairs, so it does not show whether a growing network sustains entanglement delivery over long distances.
This motivates an objective that combines the amount of delivered entanglement with the physical separation over which it is delivered, while retaining an explicit requirement on the quality of the delivered states.

The transport-capacity viewpoint introduced for classical wireless networks by Gupta and Kumar weights delivered information by the distance over which it is carried, measured in bit\(\cdot\)meters per second~\cite{gupta2000capacity}.
Their scaling laws, however, do not transfer directly to quantum networks, because they model nodes that share a wireless channel within a region of fixed area, where interference limits concurrent transmissions.
A quantum network instead distributes entanglement over a graph of link opportunities, where the topology, the aggregation of stored elementary pairs, and implementation constraints enter the scaling, and where spatial expansion and node densification must be distinguished.
We therefore adopt only the distance-weighted objective and define \emph{entanglement transport} by weighting each delivered Bell pair whose fidelity with the target Bell state meets a fixed threshold, by the physical separation of its endpoints.
Counting each fidelity-qualified Bell pair as one ebit, entanglement transport is measured in \([\mathrm{ebit}\cdot\mathrm{meter}]\), and its rates are expressed per activation slot in the resource layer and per second in the physical layer.
This objective provides a common quantity for studying how entanglement delivery scales jointly with network size, the number of active users, and communication distance, without prescribing a particular routing or physical implementation.

We therefore formulate entanglement transport through two complementary layers.
The resource layer abstracts elementary entanglement supply and measures the largest expected entanglement transport per activation slot over admissible strategies.
An activation slot is a resource-accounting round in which every link opportunity is attempted once, and elementary pairs supplied in different slots may be combined within an aggregation window.
The physical layer incorporates implementation constraints, such as decoherence, noise, storage limits, and signaling delays in wall-clock time, and measures the largest long-run expected entanglement transport per second over admissible physical implementations. 
It imposes no aggregation window, since elementary-pair retention and reset are governed by the implementation itself.
When a physical implementation is matched to the resource model and uses a bounded number of activation slots per second, the asymptotic resource-layer capacity induces a ceiling on its physical-layer service, whereas achievable service must be established for each specified implementation.

\subsection{Relation to prior work}
\label{subsec:prior-work}

To position the entanglement transport framework, we review prior work on capacity and utility, routing and storage, spatial scaling, and physical implementations.

Quantum-network capacity analyses bound entanglement and key distribution rates~\cite{pirandola2019endtoend} and include weighted multi-pair objectives~\cite{bauml2020linear}.
Utility frameworks evaluate task-dependent objectives and can incorporate rate--fidelity tradeoffs~\cite{lee2024quantum,vardoyan2022quantum}, while Azuma and Kato distinguish rates per total channel use and per unit time~\cite{azuma2017aggregating}.
Routing studies address probabilistic links and heralded information~\cite{pant2019routing}, weighted throughput over processing windows~\cite{li2020effective}, and multi-demand rate optimization under fidelity-derived path constraints~\cite{chakraborty2020entanglement}.

The quantum overlay model combines weighted rate optimization, storage across intervals, and end-to-end fidelity requirements using path flows and average purification costs~\cite{pouryousef2022overlay}.
Gatti et al.\ enforce per-request fidelity thresholds in slotted online routing with a segment-local expected-goodput metric and report a fidelity-qualified service radius in physical distance~\cite{gatti2026fidelity}.
Spatial studies average delivered entanglement or two-user capacities over endpoint pairs and graph realizations~\cite{lapeyre2012distribution,zhuang2021quantum} and analyze traffic and memory overhead as end-user populations grow in fixed or expanding regions~\cite{choi2023scalable}.
Link-layer protocols specify entanglement services with fidelity requirements~\cite{dahlberg2019link}.
Repeater studies analyze waiting times and dephasing with imperfect memories~\cite{kamin2023exact}, classical-communication effects on waiting and memory aging~\cite{haldar2025reducing}, and time-multiplexed generation and purification in fixed-duration sessions~\cite{kinos2025complete}.

Entanglement and concurrence percolation studies establish connectivity thresholds for pure-state elementary resources~\cite{acin2007entanglement,perseguers2010multipartite,meng2021concurrence,nath2026general}.
In the resource layer, they correspond to lattices of deterministically supplied pure elementary pairs with \(F_\star=1\), and each operationally specified construction, such as singlet conversion followed by swapping~\cite{acin2007entanglement} or multipartite measurements~\cite{perseguers2010multipartite}, is an admissible strategy whose achieved entanglement transport is a constructive lower bound on capacity.
Their thresholds, however, are neither weighted by distance nor normalized by the elementary pairs and activation slots that a construction consumes.
Section~\ref{sec:classification} fixes this accounting and derives entanglement transport orders for the specified operational realizations.

These studies use different objectives, resource assumptions, and normalizations, which must be aligned when comparing network scaling.

\subsection{Contributions}
The entanglement transport framework is intended as a common basis for deriving and comparing such scaling laws.
We demonstrate it on percolation-family constructions on an expanding honeycomb lattice and on homogeneous chains, where the same objective separates the effects of processing strategy, aggregation, and implementation constraints.
\begin{enumerate}
\item \textbf{Framework and matched comparison.}
We define entanglement transport capacity per activation slot and service per second for the same fidelity-qualified, distance-weighted delivery objective.
We distinguish single-slot, fixed finite-window, and asymptotic aggregation and establish their capacity ordering.
The matched comparison of two layers preserves the network, demand, joint output states, and fidelity qualification.
Under the matched comparison and the stated bound on resource-layer activation slots per second, an upper bound on asymptotic resource-layer capacity induces a physical-layer service ceiling (Section~\ref{sec:framework-definitions}).

\item \textbf{Percolation-family comparison.}
We place entanglement and concurrence percolation constructions~\cite{acin2007entanglement,perseguers2010multipartite,meng2021concurrence,nath2026general} in the resource layer on the specified expanding honeycomb lattice with one uniformly random session and common elementary-pair supply.
For the operational fixed-window constructions, namely one- and two-copy classical entanglement percolation and the two-copy construction of Acín et al.~\cite{acin2007entanglement} and the GHZ construction of Perseguers et al.~\cite{perseguers2010multipartite}, we derive achieved entanglement transport scalings strictly below and above their thresholds.
For a specified direct all-shortest-path realization of general concurrence percolation~\cite{nath2026general}, we quantify how its growing aggregation cost suppresses achieved entanglement transport, whereas concurrence percolation theory~\cite{meng2021concurrence} is compared only by its threshold because its higher-order connectivity rules may not be LOCC-realizable (Section~\ref{sec:classification}).

\item \textbf{Chain network resource-layer benchmark.}
For homogeneous chains with \(N\) nodes and \(M\) active users under random exclusive pairing, we derive expansion and densification capacity scalings using constructive and converse bounds.
Asymptotic capacity exceeds fixed-window capacity by a factor \(\Theta(N^2/M)\) for fixed link multiplicity and fixed opportunity probability strictly between zero and one (Section~\ref{sec:chain-resource}).

\item \textbf{Chain network physical-layer benchmark.}
In expansion at fixed physical parameters, the specified noisy implementation and its purification-aided variant achieve \(\Theta(M/N)\) service with different coefficients, matching the fixed-window resource-layer scaling.
The ideal-retention implementation achieves \(\Theta(N/\log N)\), a factor \(\Theta(\log N)\) below the per-second asymptotic-aggregation ceiling (Section~\ref{sec:chain-physical}).
\end{enumerate}

\section{The framework}
\label{sec:framework-definitions}

\begin{figure*}[htbp]
\centering
\resizebox{0.95\textwidth}{!}{\input{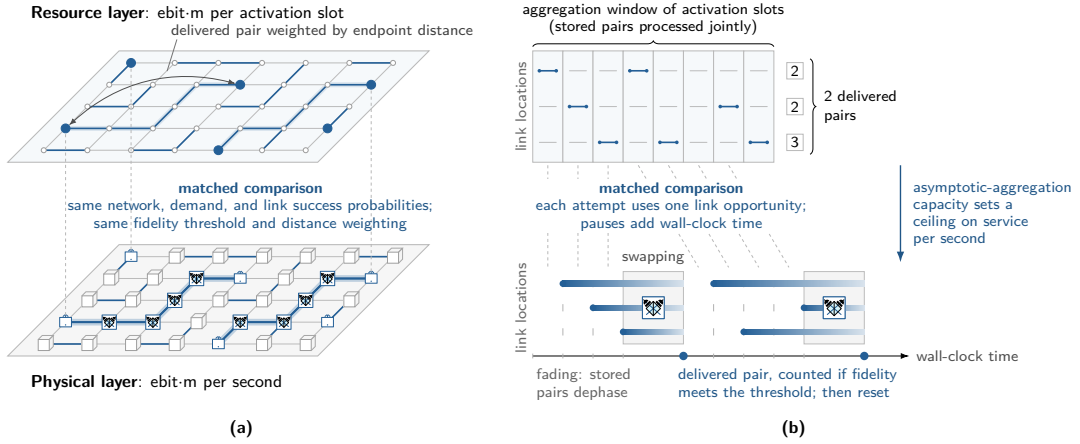}}
\caption{Two-layer entanglement transport framework.
\textbf{(a)}~The resource layer measures entanglement transport capacity per activation slot, and the physical layer measures entanglement transport service per second.
Both layers share the network, the demand, the link success probabilities, the fidelity threshold, and the distance weighting, under which each delivered pair meeting the fidelity threshold is weighted by its endpoint distance.
\textbf{(b)}~The two layers differ in time accounting, shown schematically for three link locations along one route, with identical link outcomes drawn in both layers for illustration.
In the resource layer, elementary pairs supplied within an aggregation window of activation slots are stored and jointly processed.
In the physical layer, generation attempts, heralds, memory dephasing, and swapping take place in wall-clock time, following the noisy implementation of Section~\ref{sec:chain-physical}, with durations not to scale.
Under the matched comparison of Section~\ref{subsec:cross-layer} and a bound on resource-layer activation slots per second, asymptotic-aggregation capacity sets a ceiling on physical-layer service, whereas achievable service must be established for each implementation.}
\label{fig:two-layer}
\end{figure*}

Figure~\ref{fig:two-layer} summarizes the framework: the resource layer describes entanglement transport capacity per activation slot, and the physical layer describes entanglement transport service per second under explicit implementation constraints.
Figure~\ref{fig:chain-model} illustrates the resource-layer elements on a six-node chain, which serves throughout as a running example.

\begin{figure}[htbp]
\centering
\resizebox{0.75\textwidth}{!}{
%
\definecolor{sessionblue}{HTML}{0072B2}%
\definecolor{sessionorange}{HTML}{D55E00}%
\definecolor{linkdark}{HTML}{3A3A3A}%
\definecolor{linklight}{HTML}{C4C4C4}%
\definecolor{nodelight}{HTML}{F4F4F4}%
\definecolor{tagcol}{HTML}{6E6E6E}%
\tikzset{%
  every node/.style={font=\small},%
  network node/.style={circle,draw=linkdark,fill=white,line width=0.55pt,minimum size=4.0mm,inner sep=0pt},%
  inactive node/.style={network node,fill=nodelight},%
  demand/.style={densely dashed,line width=0.75pt},%
  panel title/.style={font=\scriptsize\bfseries,anchor=west},%
  legend text/.style={font=\scriptsize,anchor=west},%
  note/.style={font=\scriptsize,anchor=west},%
  realized/.style={line width=1.0pt,draw=linkdark},%
  standing/.style={line width=1.0pt,draw=linklight},%
  dimline/.style={line width=0.4pt,draw=linkdark!70,{Latex[length=1.3mm,width=1.0mm]}-{Latex[length=1.3mm,width=1.0mm]}},%
  ledger/.style={draw=linkdark!60,line width=0.45pt,fill=nodelight,rounded corners=0.6pt,minimum width=4.6mm,minimum height=3.6mm,inner sep=1pt,font=\scriptsize}%
}%
\newcommand{\nodesep}{1.00}%
%
\newcommand{\bundlew}[5]{%
  \foreach \kk in {1,2}{%
    \ifnum\kk>#3
      \draw[standing] ($(#1)+(0,{(\kk-1.5)*0.22})$) -- ($(#2)+(0,{(\kk-1.5)*0.22})$);%
    \else
      \ifnum\kk>#4
        \draw[realized] ($(#1)+(0,{(\kk-1.5)*0.22})$) -- ($(#2)+(0,{(\kk-1.5)*0.22})$);%
      \else
        \draw[line width=1.5pt,draw=#5] ($(#1)+(0,{(\kk-1.5)*0.22})$) -- ($(#2)+(0,{(\kk-1.5)*0.22})$);%
      \fi
    \fi
  }%
}%
\newcommand{\chaincoords}[1]{%
  \foreach \ii in {0,...,5}{\coordinate (#1-\ii) at ({\nodesep*\ii},0);}%
}%
\newcommand{\chainstanding}[1]{%
  \foreach \ii [evaluate=\ii as \nx using int(\ii+1)] in {0,...,4}{\bundlew{#1-\ii}{#1-\nx}{0}{0}{linkdark}}%
}%
\newcommand{\chainnodes}[1]{%
  \foreach \ii in {0,...,5}{%
    \ifnum\ii=0 \node[network node,draw=sessionblue,line width=1pt] at (#1-\ii) {};%
    \else\ifnum\ii=5 \node[network node,draw=sessionblue,line width=1pt] at (#1-\ii) {};%
    \else\ifnum\ii=2 \node[network node,draw=sessionorange,line width=1pt] at (#1-\ii) {};%
    \else\ifnum\ii=4 \node[network node,draw=sessionorange,line width=1pt] at (#1-\ii) {};%
    \else \node[inactive node] at (#1-\ii) {};%
    \fi\fi\fi\fi
    \node[anchor=north,font=\scriptsize] at ($(#1-\ii)+(0,-0.19)$) {\(\ii\)};%
  }%
}%
\newcommand{\chaindemand}[1]{%
  \draw[demand,sessionblue] ($(#1-0)+(0,0.11)$) .. controls ($(#1-0)+(1.45,1.05)$) and ($(#1-5)+(-1.45,1.05)$) .. ($(#1-5)+(0,0.11)$);%
  \draw[demand,sessionorange] ($(#1-2)+(0,0.11)$) .. controls ($(#1-2)+(0.60,0.62)$) and ($(#1-4)+(-0.60,0.62)$) .. ($(#1-4)+(0,0.11)$);%
}%
%
\newcommand{\pairglyph}[3]{%
  \draw[line width=0.9pt,draw=#3,decorate,decoration={snake,amplitude=0.55pt,segment length=4.2pt}] (#1) -- (#2);%
  \fill[#3] (#1) circle (1.5pt);%
  \fill[#3] (#2) circle (1.5pt);%
}%
\begin{tikzpicture}

\begin{scope}[shift={(2.50,10.30)}]
  \node[panel title,anchor=south] at (2.50,1.30) {(a) Resource graph \(G=(V,E)\) and demand \(\mathcal D\)};
  \chaincoords{a}
  \chainstanding{a}
  \chaindemand{a}
  \chainnodes{a}
  \draw[dimline] ($(a-0)+(0,-0.56)$) -- ($(a-1)+(0,-0.56)$);
  \node[anchor=north,font=\scriptsize] at ($(a-0)!0.5!(a-1)+(0,-0.58)$) {\(a_N\)};
  \draw[decorate,decoration={brace,amplitude=3pt,mirror},draw=linkdark!70,line width=0.4pt]
       ($(a-3)+(0.12,-0.52)$) -- ($(a-4)+(-0.12,-0.52)$);
  \node[anchor=north,font=\scriptsize,text=linkdark] at ($(a-3)!0.5!(a-4)+(0,-0.64)$) {\(w=2\)};
\end{scope}

\begin{scope}[shift={(2.50,7.22)}]
  \node[panel title,anchor=south] at (2.50,1.25) {(b) Slot realization \(G_1=(V,E_1)\)};
  \chaincoords{b}
  \bundlew{b-0}{b-1}{1}{1}{sessionblue}
  \bundlew{b-1}{b-2}{2}{1}{sessionblue}
  \bundlew{b-2}{b-3}{1}{1}{sessionblue}
  \bundlew{b-3}{b-4}{1}{1}{sessionblue}
  \bundlew{b-4}{b-5}{2}{1}{sessionblue}
  \chainnodes{b}
  \coordinate (bA) at ($(b-0)+(0,0.60)$);
  \coordinate (bB) at ($(b-5)+(0,0.60)$);
  \draw[dotted,line width=0.5pt,draw=sessionblue] ($(b-0)+(0,0.22)$) -- (bA);
  \draw[dotted,line width=0.5pt,draw=sessionblue] ($(b-5)+(0,0.22)$) -- (bB);
  \pairglyph{bA}{bB}{sessionblue}
  \node[font=\scriptsize,text=sessionblue,anchor=south] at ($(bA)!0.5!(bB)+(0,0.03)$) {\(\rho_{0,5}\)};
  \node[anchor=north,font=\scriptsize] at ({2.5*\nodesep},-0.62) {entanglement transport \(5a_N\)};
\end{scope}

\begin{scope}[shift={(2.50,4.16)}]
  \node[panel title,anchor=south] at (2.50,1.25) {(c) Slot realization \(G_2=(V,E_2)\)};
  \chaincoords{c}
  \bundlew{c-0}{c-1}{1}{0}{sessionorange}
  \bundlew{c-1}{c-2}{0}{0}{sessionorange}
  \bundlew{c-2}{c-3}{2}{1}{sessionorange}
  \bundlew{c-3}{c-4}{1}{1}{sessionorange}
  \bundlew{c-4}{c-5}{1}{0}{sessionorange}
  \chainnodes{c}
  \coordinate (cA) at ($(c-2)+(0,0.60)$);
  \coordinate (cB) at ($(c-4)+(0,0.60)$);
  \draw[dotted,line width=0.5pt,draw=sessionorange] ($(c-2)+(0,0.22)$) -- (cA);
  \draw[dotted,line width=0.5pt,draw=sessionorange] ($(c-4)+(0,0.22)$) -- (cB);
  \pairglyph{cA}{cB}{sessionorange}
  \node[font=\scriptsize,text=sessionorange,anchor=south] at ($(cA)!0.5!(cB)+(0,0.03)$) {\(\rho_{2,4}\)};
  \node[anchor=north,font=\scriptsize] at ({2.5*\nodesep},-0.62) {entanglement transport \(2a_N\)};
\end{scope}

\begin{scope}[shift={(2.50,0.55)}]
  \node[panel title,anchor=south] at (2.50,1.80) {(d) Aggregation over a window of \(W=2\) activation slots};
  \chaincoords{d}
  \chainstanding{d}
  \chainnodes{d}
  \coordinate (dA) at ($(d-0)+(0,0.60)$);
  \coordinate (dB) at ($(d-5)+(0,0.60)$);
  \coordinate (dC) at ($(d-0)+(0,1.20)$);
  \coordinate (dD) at ($(d-5)+(0,1.20)$);
  \draw[dotted,line width=0.5pt,draw=sessionblue] ($(d-0)+(0,0.22)$) -- (dC);
  \draw[dotted,line width=0.5pt,draw=sessionblue] ($(d-5)+(0,0.22)$) -- (dD);
  \pairglyph{dA}{dB}{sessionblue}
  \pairglyph{dC}{dD}{sessionblue}
  \node[font=\scriptsize,text=sessionblue,anchor=south] at ($(dA)!0.5!(dB)+(0,0.03)$) {\(\rho_{0,5}\)};
  \node[font=\scriptsize,text=sessionblue,anchor=south] at ($(dC)!0.5!(dD)+(0,0.03)$) {\(\rho_{0,5}\)};
  \foreach \ii/\val in {0/2,1/2,2/3,3/2,4/3}{
    \node[ledger] at ({\nodesep*\ii+0.5*\nodesep},-0.72) {\val};}
  \node[anchor=north,font=\scriptsize] at ({2.5*\nodesep},-1.05) {entanglement transport \(5a_N\) per activation slot};
\end{scope}

\begin{scope}
  \draw[draw=linkdark!25,line width=0.4pt] (0.00,-1.27) -- (10.00,-1.27);
  \draw[standing] (0.00,-1.60) -- (0.55,-1.60);
  \node[legend text] at (0.65,-1.60) {link opportunity in \(E\)};
  \draw[realized] (5.00,-1.60) -- (5.55,-1.60);
  \node[legend text] at (5.65,-1.60) {realized link in \(E_\tau\)};
  \draw[line width=1.5pt,draw=sessionblue] (0.00,-2.13) -- (0.55,-2.13);
  \node[legend text] at (0.65,-2.13) {realized link used by a delivery};
  \coordinate (lgA) at (5.05,-2.13);
  \coordinate (lgB) at (5.50,-2.13);
  \pairglyph{lgA}{lgB}{sessionblue}
  \node[legend text] at (5.65,-2.13) {delivered Bell pair};
  \draw[demand,sessionblue] (0.00,-2.66) -- (0.55,-2.66);
  \node[legend text] at (0.65,-2.66) {session in \(\mathcal D\)};
  \node[inactive node] at (5.28,-2.66) {};
  \node[legend text] at (5.65,-2.66) {node in \(V\); ringed if an active user};
\end{scope}

\end{tikzpicture}%
}
\caption{A resource-layer chain network.
(a) shows the resource graph with \(N=6\), in which each of the five link locations carries \(w=2\) link opportunities. 
Link opportunities are attempted in every activation slot and succeed independently with probability \(p\), with each realized link supplying one ideal Bell pair.
Over the active users \(U=\{0,2,4,5\}\), with \(M=4\), the demand \(\mathcal D=\{(0,5),(2,4)\}\) is announced network-wide before the first activation slot and is held fixed throughout the aggregation window.
A delivered state is counted only if \(F(\rho_{st},\Phi^+)\ge F_\star\), and each counted pair is weighted by the endpoint separation \(d(s_i,t_i)=a_Nh_i\). 
\(C_{\mathrm{ET}}^{(W)}\) is the largest expected entanglement transport per activation slot over admissible strategies.
(b) and (c) show two slot realizations, with the realized links consumed by a delivery drawn in the color of the session they serve.
In (b), link locations \((2,3)\) and \((3,4)\) each carry a single realized link, so the two sessions compete for the same resources and cannot both be served. 
A strategy maximizing entanglement transport serves \((0,5)\), yielding entanglement transport \(5a_N\), rather than \((2,4)\), which would yield \(2a_N\).
In (c), no link is realized at link location \((1,2)\), so session \((0,5)\) has no route and only \((2,4)\) is served.
(d) aggregates the two activation slots under ideal resource storage and joint processing, with the boxed integers giving the realized links accumulated at each link location over the window.
The accumulated realized links deliver two states for session \((0,5)\), giving entanglement transport \(10a_N\), or \(5a_N\) per activation slot, compared with \(3.5a_N\) per activation slot when the two slots are counted separately.
}
\label{fig:chain-model}
\end{figure}

\subsection{Networks, demand, and growth}
\label{subsec:networks}

\paragraph{Resource graph, link opportunity probability, and activation slot.}
The resource graph \(G=(V,E)\) is a multigraph without self-loops.
The set \(V\) collects the communicating or supporting \textit{network nodes}, and \(N=|V|\) denotes the infrastructure size.
The set \(E\) collects the \textit{link opportunities} between their endpoints.
The \(w\) parallel link opportunities at one \textit{link location} represent bounded \textit{link multiplicity}.
An \textit{activation slot} is the resource-accounting unit in which every link opportunity is attempted once, rather than a fixed amount of wall-clock time.
The attempt at link opportunity \(e\) succeeds with \textit{link opportunity probability} \(p_e\), where independent link opportunity probability is the baseline model.
In activation slot \(\tau\), the link opportunities whose attempts succeed are the \textit{realized links}.
They form the \textit{slot realization}, a spanning sub-multigraph of \(G\),
\begin{align}
G_\tau=(V,E_\tau),\qquad E_\tau\subseteq E.
\end{align}
Each realized link supplies one \textit{elementary pair} in a state declared by the resource layer.
How realized links may be used is specified by the admissible strategies of Sec.~\ref{subsec:aggregation}.

\textit{Chain example.}
Here \(V=\{0,\ldots,N-1\}\), each adjacent-node link location carries \(w\) parallel link opportunities with \(w\) fixed independently of \(N\), every link opportunity has \(p_e=p\), and each realized link supplies an ideal Bell pair.

\paragraph{Active users and communication demand.}
Let \(U\subseteq V\) be the active-user set, with \(M=|U|\) the active-user population.
The scales \(N\) and \(M\) may grow independently, subject to \(M\le N\).
The baseline demand model is \textit{random exclusive pairing}.
For even \(M\), the set \(U\) is sampled uniformly at random among all \(M\)-subsets of \(V\), and the members of \(U\) are then paired uniformly and exclusively at random into \(M/2\) endpoint-disjoint \textit{sessions} \(\mathcal D=\{(s_i,t_i)\}_{i=1}^{M/2}\), where session \(i\) demands entanglement between endpoints \(s_i\) and \(t_i\).
The active-user set and pairing are sampled once for each \textit{aggregation window}, a fixed number of consecutive activation slots formalized in Sec.~\ref{subsec:aggregation}.
\(\mathcal D\) is announced network-wide before any link opportunity is attempted in the first activation slot and is held fixed throughout the aggregation window.
Resource-layer expectations average over (i) the demand sampling, (ii) the link opportunity outcomes, and (iii) any strategy randomization.
Additional physical randomness is included in physical-layer expectations as appropriate.

\textit{Chain example.}
In Fig.~\ref{fig:chain-model}, \(U=\{0,2,4,5\}\) and \(\mathcal D=\{(0,5),(2,4)\}\), so \(M=4\) active users form two sessions.

\paragraph{Network growth and physical geometry.}
A network family must specify how both its resource graph and its physical geometry change with \(N\).
Two baseline growth regimes are distinguished.
In \textit{spatial expansion} the covered region of the deployment space grows with \(N\), typically at fixed node density.
In \textit{node densification} the covered region remains fixed while the number of nodes increases.
For session \(i\), \(h_i\) denotes the dimensionless shortest-path hop distance between \(s_i\) and \(t_i\) in \(G\), and \(d(s_i,t_i)\) denotes their physical endpoint separation in meters.
In general topologies, \(h_i\) and \(d(s_i,t_i)\) need not be proportional. 

\textit{Chain example.}
With \(a_N\) the spacing between consecutive nodes,
\[
a_N=
\begin{cases}
a, & \text{spatial expansion},\\[1mm]
L/(N-1), & \text{densification over fixed span \(L\)},
\end{cases}
\]
so expansion increases the physical span with \(N\) while densification decreases the spacing within a fixed span, and \(h_i=|t_i-s_i|\) with \(d(s_i,t_i)=a_N h_i\).

\subsection{Entanglement transport}
Let \(\rho_{st}\) denote the final two-qubit state of a delivered pair for session \((s,t)\), with one qubit at each endpoint, after all permitted processing.
The state is conditioned on the realized links and on the classical record identifying the output and its session, while unobserved noise is averaged over.
If several outputs are counted jointly, their two-qubit states are marginals of the joint output state.
The pair is counted only if its fidelity with the target Bell state satisfies \(F(\rho_{st},\Phi^+)=\bra{\Phi^+}\rho_{st}\ket{\Phi^+}\ge F_\star\), where \(\ket{\Phi^+}=(\ket{00}+\ket{11})/\sqrt{2}\) and \(F_\star\in(1/2,1]\) is fixed.
Converse bounds derived through the fidelity calibration below additionally require \(F_\star\in(F_{\min},1)\), with \(F_{\min}\) defined below Eq.~\eqref{eq:fidelity-calibration-factor}.
A delivered pair satisfying this condition is called a \textit{fidelity-qualified Bell pair}, which is not necessarily an ideal pure ebit.
The qualification target is \(\Phi^+\) irrespective of the elementary pair state. 
We define \textit{entanglement transport}, measured in \([\mathrm{ebit}\cdot\mathrm{meter}]\), as the sum of the physical endpoint separations of all fidelity-qualified Bell pairs delivered over a specified interval.

\textit{Chain example.}
One fidelity-qualified Bell pair delivered for session \((0,4)\) contributes \(4a_N\) ebit\(\cdot\)meters of entanglement transport.

\subsection{Aggregation regimes}
\label{subsec:aggregation}

\paragraph{Aggregation window and admissible strategies.}
An \textit{aggregation window} of \(W\) consecutive activation slots specifies how many slots may contribute realized links before entanglement transport is counted.
As a baseline model, the resource layer permits ideal storage across the window when \(W>1\).
Finite usable memory lifetimes motivate restricting accumulation across activation slots: the window limits which activation slots may contribute elementary pairs that are stored and jointly processed~\cite{dhara2021subexponential}.
The window limits accumulation to a common block of slots, unlike a memory-cutoff policy that assigns each stored pair a maximum age~\cite{khatri2021policies}.
In the \textit{single-slot regime}, \(W=1\) and output is counted separately in each activation slot.
In a \textit{finite-window regime}, resources may be stored and jointly processed over \(1<W<\infty\) activation slots, with \(W\) fixed independently of \(N\).
In the \textit{asymptotic aggregation regime}, arbitrarily many activation slots may be combined under the resource layer's ideal storage and joint-processing abstraction.
Aggregation windows start empty, so only resources supplied within the window may contribute to its entanglement transport, and the demand remains fixed throughout.
An \textit{admissible strategy} specifies the permitted routing, allocation, swapping, and adaptive local operations and classical communication (LOCC), together with the heralded realized link information on which these operations may depend.
A \textit{pre-heralding} strategy may use \(\mathcal D\) but not the current-slot link outcomes when selecting its schedule, whereas a \textit{post-heralding} strategy may adapt to those outcomes.
In the resource layer, classical communication does not consume additional activation slots, so information assumptions enter only through the pre- or post-heralding restriction.
In the physical layer, classical communication incurs signaling delays in wall-clock time, and each implementation specifies which delays it includes.

\paragraph{Entanglement transport capacity.}
Let \(B_i^{(W)}\) denote the number of Bell pairs delivered for session \(i\) over the \(W\)-slot aggregation window that satisfy the fidelity threshold \(F_\star\).
The network-wide resource-layer entanglement transport capacity is
\begin{equation}
C_{\mathrm{ET}}^{(W)}(N,M;F_\star) =
\sup_{\substack{\text{admissible}\\\text{strategies}}}
\frac{1}{W} \mathbb{E}\!\left[ \sum_{i=1}^{|\mathcal D|} d(s_i,t_i)B_i^{(W)} \right]
\quad
[\mathrm{ebit}\cdot\mathrm{meter}/\mathrm{slot}].
\label{eq:resource-et-capacity}
\end{equation}
Because capacity is normalized per activation slot, a strategy that combines elementary pairs from several slots is charged for that additional supply, so constructions with different storage requirements, such as the \(W=1\) and \(W=2\) percolation constructions of Sec.~\ref{sec:classification}, are compared under common resource counting.
The arguments \(N\), \(M\), and \(F_\star\) are suppressed when clear. 
\(F_\star\) is retained when capacities at different fidelity thresholds are compared.
The perfect-output capacity is \(C_{\mathrm{ET}}^{(W)}(N,M;1)\).

Its session-normalized counterpart is
\begin{equation}
c_{\mathrm{ET}}^{(W)}(N,M;F_\star) := \frac{C_{\mathrm{ET}}^{(W)}(N,M;F_\star)}{|\mathcal D|}
\quad
[\mathrm{ebit}\cdot\mathrm{meter}/\mathrm{slot}/\mathrm{session}].
\label{eq:session-normalized-capacity}
\end{equation}
This normalization measures average transport per requested session without imposing equal service or max--min fairness.
The asymptotic aggregation capacity is
\begin{equation}
C_{\mathrm{ET}}^{(\infty)}(N,M;F_\star) := \lim_{W\to\infty}C_{\mathrm{ET}}^{(W)}(N,M;F_\star).
\label{eq:asymptotic-capacity}
\end{equation}

\paragraph{Aggregation ordering.}
Fix \(N\), \(M\), and \(F_\star\), and consider the total entanglement transport over a window of \(W\) activation slots, \(W\,C_{\mathrm{ET}}^{(W)}\).
Three baseline properties are used: the link opportunity outcomes of different activation slots are independent, every window starts empty with the announced demand fixed throughout, and the admissible-strategy class is closed under concatenation.
Any admissible \(k\)-slot strategy and any admissible \(\ell\)-slot strategy can therefore be run consecutively, with unused resources discarded at the boundary, and the expected total entanglement transport of the concatenated strategy is the sum of the two contributions.
Taking the supremum over all such concatenations gives
\[
(k+\ell)\,C_{\mathrm{ET}}^{(k+\ell)}
\ge
k\,C_{\mathrm{ET}}^{(k)}
+
\ell\,C_{\mathrm{ET}}^{(\ell)},
\]
so \(W\,C_{\mathrm{ET}}^{(W)}\) is superadditive.
Fekete's lemma therefore gives
\begin{equation}
C_{\mathrm{ET}}^{(\infty)} = \sup_{W\ge1}C_{\mathrm{ET}}^{(W)},
\qquad
C_{\mathrm{ET}}^{(1)} \le C_{\mathrm{ET}}^{(W)} \le C_{\mathrm{ET}}^{(\infty)},
\label{eq:aggregation-ordering}
\end{equation}
and the limit in Eq.~\eqref{eq:asymptotic-capacity} exists and equals this supremum.
This ordering brackets the contribution of aggregation across activation slots: for the homogeneous chain, asymptotic aggregation exceeds fixed-window capacity by a factor \(\Theta(N^2/M)\) (Sec.~\ref{sec:chain-resource}), and in Fig.~\ref{fig:chain-model}(d) aggregating two slot realizations raises the entanglement transport from \(3.5a_N\) to \(5a_N\) per activation slot.
We use \(\underline C_{\mathrm{ET}}^{(1)}\le C_{\mathrm{ET}}^{(1)}\) for a constructive lower bound and \(\overline C_{\mathrm{ET}}^{(\infty)}\ge C_{\mathrm{ET}}^{(\infty)}\) for a converse upper bound, both under the same fidelity-qualified transport definition.

A converse starts from an upper bound on the squashed entanglement~\cite{christandl2004squashed} of the elementary pairs supplied across a partition of the network nodes into two sets, where each ideal Bell pair carries one unit.
Because resource-layer LOCC cannot increase squashed entanglement across the partition, the joint state of the fidelity-qualified Bell pairs with endpoints in different sets carries on average at most this supplied amount.
A fidelity-qualified Bell pair may carry less than one unit, so this amount does not by itself bound their number, and we therefore define the \textit{fidelity calibration factor}
\begin{equation}
\kappa(F_\star)
:=
\left[
1+F_\star\log_2F_\star+(1-F_\star)\log_2\!\left(\frac{1-F_\star}{3}\right)
\right]^{-1}.
\label{eq:fidelity-calibration-factor}
\end{equation}
This factor is defined for \(F_\star\in(F_{\min},1)\), where \(F_{\min}\simeq0.8107\) is the unique root in \((0,1)\) of the denominator.
Twirling and the hashing bound show that each fidelity-qualified Bell pair carries at least \(1/\kappa(F_\star)\) units of squashed entanglement, so multiplying the upper bound on the supplied squashed entanglement by \(\kappa(F_\star)\) gives a converse, not necessarily tight, for their expected number, including for correlated outputs (Appendix~\ref{app:fidelity-calibration})~\cite{bennett1996mixed}.
For fixed \(F_\star\), \(\kappa(F_\star)\) is independent of \(N,M\), so this calibration preserves the bound's scaling exponents.

\subsection{Physical layer and matched comparison}
\label{subsec:cross-layer}

\paragraph{Physical layer.}
The physical layer specifies implementation constraints and hardware imperfections, including decoherence, noise, storage limits, and signaling delays in wall-clock time.
An \textit{admissible physical implementation} is any protocol obeying these constraints.
Physical service is not assigned an aggregation window \(W\); elementary pair retention and reset are governed by the physical implementation.

\paragraph{Matched comparison.}
A matched comparison translates the resource layer to the physical layer without changing the resource graph \(G\), demand \(\mathcal D\), fidelity threshold \(F_\star\), or distance weighting \(d\).
For each link opportunity \(e\), the corresponding physical generation attempt has the same link opportunity probability \(p_e\) and the same independence assumptions as in the resource layer.
Each physical generation attempt is represented by a distinct corresponding resource-layer opportunity, without conditioning that assignment on an unrevealed link outcome, and the physical-layer elementary pair must be obtainable from the resource-layer elementary pair through local operations at its two endpoints.
Under this correspondence, every finite physical execution relevant to the counted outputs must be reproducible by an admissible resource-layer strategy, including the joint output state and classical record used for fidelity qualification.

\textit{Chain example.}
Consider a physical implementation of the chain in Fig.~\ref{fig:chain-model} that keeps its resource graph, demand, link opportunity probability \(p\), fidelity threshold \(F_\star\), and distance weighting.
If each link location makes \(w\) independent generation attempts per round and the attempts of round \(\tau\) are assigned to the link opportunities of activation slot \(\tau\) before their outcomes are revealed, each physical attempt corresponds to a distinct resource-layer opportunity.
Hardware noise, such as initialization errors, memory decoherence, and swapping errors, acts through local operations on the elementary pairs, so the physical implementation corresponds to one particular resource-layer strategy that applies the same noise to ideal Bell pairs.
By contrast, attempts that succeed with probability above \(p\) or with correlated outcomes are not matched.

\paragraph{Entanglement transport service.}
For physical service, \(\mathcal D\) is sampled once from the same demand law and held fixed as \(T\to\infty\), with expectations averaging over the demand sampling randomness.
Let \(B_i(T)\) be the number of fidelity-qualified Bell pairs delivered for session \(i\) during a wall-clock interval of \(T\) seconds under an admissible physical implementation.
The network-wide physical-layer entanglement transport service is
\begin{equation}
S_{\mathrm{ET}}(N,M;F_\star) = \sup_{\substack{\text{admissible}\\\text{implementations}}}
\limsup_{T\to\infty} \frac{1}{T} \mathbb{E} \left[ \sum_{i=1}^{|\mathcal D|}d(s_i,t_i)B_i(T) \right]
\quad [\mathrm{ebit}\cdot\mathrm{meter}/\mathrm{second}].
\label{eq:physical-et-service}
\end{equation}
The limit superior is used because admissible implementations are not assumed to have convergent long-run averages.
For a specified admissible implementation,
\[
\underline S_{\mathrm{ET}}(N,M;F_\star) = \limsup_{T\to\infty} \frac{1}{T} \mathbb{E}\!\left[ \sum_{i=1}^{|\mathcal D|} d(s_i,t_i)B_i(T) \right]
\]
is its achieved service.
Thus \(\underline S_{\mathrm{ET}}(N,M;F_\star)\le S_{\mathrm{ET}}(N,M;F_\star)\).

\paragraph{Single-slot reference and physical-service ceiling.}
For a matched comparison, let \(r\,[\mathrm{slot}/\mathrm{s}]\) denote the \textit{reference slot rate}, with reference interval \(1/r\), used to express the single-slot resource capacity on a per-second scale.
Protocol-dependent generation pauses, processing, feedback, waiting, and other delays remain part of the wall-clock service rather than being absorbed into \(r\).
For a specified implementation and \(0<C_{\mathrm{ET}}^{(1)}(N,M;F_\star)<\infty\), define the achieved \textit{single-slot reference ratio} as
\begin{equation}
\frac{\underline S_{\mathrm{ET}}(N,M;F_\star)}{r\,C_{\mathrm{ET}}^{(1)}(N,M;F_\star)}.
\label{eq:single-slot-reference-ratio}
\end{equation}
This ratio is a normalization against the matched \(W=1\) resource capacity.
The value one is the normalization point \(\underline S_{\mathrm{ET}}(N,M;F_\star)=rC_{\mathrm{ET}}^{(1)}(N,M;F_\star)\); physical constraints may reduce the ratio, while accumulation of elementary pairs may increase it above one.
Consider a specified implementation satisfying the matched comparison and suppose that, for every \(\epsilon>0\), the physical execution over every sufficiently long interval of duration \(T\), for every demand and protocol realization, can be represented using at most \(\lceil(r+\epsilon)T\rceil\) resource-layer activation slots.
Then its achieved service satisfies
\begin{equation}
\underline S_{\mathrm{ET}}(N,M;F_\star) \le rC_{\mathrm{ET}}^{(\infty)}(N,M;F_\star).
\label{eq:physical-service-ceiling}
\end{equation}
This follows by representing each sufficiently long physical execution by a resource-layer aggregation window whose activation-slot rate approaches \(r\) and using \(C_{\mathrm{ET}}^{(W)}\le C_{\mathrm{ET}}^{(\infty)}\).
Physical service itself has no imposed aggregation window and the resource-layer window is used only for this comparison.

\subsection{Framework extensions}
The concrete modeling choices above define the baseline instantiation studied in this paper, and the framework is not limited to the baseline.
Alternative instantiations may change the demand law, counting rule, distance functional \(d\), growth prescription, elementary pair, or the rule by which link opportunities are supplied.
Such changes preserve the distinction between resource-layer capacity per activation slot and physical-layer service per second, but the matched comparison and the resulting scaling laws must be established for the chosen instantiation.
Changing the independence assumptions requires a separate analysis of Eq.~\eqref{eq:aggregation-ordering}.
All benchmark results use the stated choices unless explicitly redefined.

\section{Percolation-family comparison}
\label{sec:classification}

\begin{table}[t]
\centering
\caption{Percolation-family constructions on the common regular fixed-aspect-ratio expanding honeycomb resource-graph family at fixed nearest-neighbor physical spacing, with \(p_e=1\), \(w=1\), \(F_\star=1\), one uniformly random session (\(M=2\)), and ideal storage for \(W>1\).
Each link location supplies one copy of \(\ket{\psi(\theta)}=\cos\theta\ket{00}+\sin\theta\ket{11}\) per activation slot, with \(0<\theta<\pi/4\) fixed independently of \(N\).
The threshold column gives the elementary-pair single-copy singlet-conversion probability (SCP) \(2\sin^2\theta_{\mathrm{th}}\) at the corresponding percolation threshold.
The classical entanglement percolation (CEP), two-copy CEP, and Acín threshold entries are decimal evaluations of exact expressions in Appendix~\ref{app:percolation-comparison}; the Perseguers GHZ, ConPT, and GCP threshold entries are source numerical results or transformations thereof.
For CEP, the Perseguers GHZ construction, the two-copy CEP, and the Acín two-copy construction, the last two columns give achieved entanglement transport per activation slot for parameters fixed strictly below or above the corresponding critical point. 
Numerical threshold entries are estimates where indicated.
Parenthesized digits give the reported uncertainty in the last digit.
No operational entanglement transport scaling is assigned to ConPT on the honeycomb lattice.
For the GCP direct all-shortest-path realization, the aggregation column gives the minimum window required by that preparation, and the last two columns give the entanglement transport achieved by independent singlet conversion followed by maximum pure-Bell-pair routing.
These orders also hold at the reported threshold.
Derivations, including bounds for alternative exact preparations of the GCP effective states, are given in Appendix~\ref{app:percolation-comparison}.
}
\label{tab:percolation-comparison}
\footnotesize
{\setlength{\tabcolsep}{5pt}
\renewcommand{\arraystretch}{1.10}
\begin{tabularx}{\textwidth}{@{}Xm{0.12\linewidth}m{0.16\linewidth}m{0.21\linewidth}m{0.21\linewidth}@{}}
\toprule
Construction
& Aggregation window \(W\)
& Elementary-pair SCP at threshold
& Entanglement transport below threshold
& Entanglement transport above threshold
\\
\midrule
CEP~\cite{acin2007entanglement}
& \(1\)
& \(0.65270\)
& \(\Theta(N^{-1})\)
& \(\Theta(\sqrt N)\)
\\[0.2em]
Perseguers GHZ~\cite{perseguers2010multipartite}
& \(1\)
& \(0.609(0)\)
& \(\Theta(N^{-1})\)
& \(\Theta(\sqrt N)\)
\\[0.2em]
two-copy CEP~\cite{acin2007entanglement}
& \(2\)
& \(0.35848\)
& \(\Theta(N^{-1})\)
& \(\Theta(\sqrt N)\)
\\[0.2em]
Acín two-copy~\cite{acin2007entanglement}
& \(2\)
& \(0.34730\)
& \(\Theta(N^{-1})\)
& \(\Theta(\sqrt N)\)
\\[0.2em]
ConPT~\cite{meng2021concurrence}
& \(1\)
& \(0.30^{+0.10}_{-0.08}\)
& not claimed
& not claimed
\\[0.2em]
GCP direct all-shortest-path realization~\cite{nath2026general}
& \(\exp[\Theta(\sqrt N)]\)
& \(0.197(5)\)
& \(\exp[-\Theta(\sqrt N)]\)
& \(\exp[-\Theta(\sqrt N)]\)
\\
\bottomrule
\end{tabularx}}
\end{table}

\paragraph{Comparison assumptions.}
We compare several percolation-family constructions on a common family of finite honeycomb resource graphs growing by spatial expansion at fixed nearest-neighbor physical spacing, with the graph shape and aspect ratio fixed as \(N\) grows.
Throughout this comparison, \(0<\theta<\pi/4\) is fixed independently of \(N\), the fidelity threshold is \(F_\star=1\), each nearest-neighbor link location has unit multiplicity \(w=1\), and each aggregation window contains one uniformly random session, so \(M=2\).
Each link location deterministically supplies one copy of the elementary pair
\begin{equation}
\ket{\psi(\theta)} = \cos\theta\ket{00} + \sin\theta\ket{11} \label{eq:perc_elementary_pair}
\end{equation}
per activation slot.
The choice \(w=1\) fixes a common resource graph but is not unique.
In particular, two elementary pairs per honeycomb link location can be supplied either over \(W=2\) activation slots with \(w=1\) and ideal storage or within one activation slot with \(w=2\); the two representations supply the same elementary pairs but differ in resources supplied per activation slot.
Because \(F_\star=1\), every pair counted by \(B_i^{(W)}\) is an ideal target Bell pair.
For the elementary-pair state in Eq.~\eqref{eq:perc_elementary_pair}, the single-copy singlet-conversion probability (SCP) is \(2\sin^2\theta\).
For CEP, the Perseguers GHZ construction, two-copy CEP, and the Acín two-copy construction, the below- and above-threshold entanglement transport statements exclude the critical point itself.
For the specified GCP direct all-shortest-path realization, the stated stretched-exponential orders also hold at its threshold.
For each construction, the comparison specifies either an aggregation window \(W\) fixed independently of \(N\) or its growth order with \(N\).
The comparison does not evaluate the asymptotic aggregation capacity \(C_{\mathrm{ET}}^{(\infty)}\).

\paragraph{Classical and quantum entanglement percolation.}
For CEP, local singlet conversion produces a pure Bell pair with probability \(2\sin^2\theta\), so the exact honeycomb bond threshold gives the elementary-pair threshold \(2\sin^2\theta_{\mathrm{th}}=1-2\sin(\pi/18)=0.65270\ldots\)~\cite{acin2007entanglement}.
The Perseguers GHZ construction starts from one elementary pair per honeycomb link location and uses a fixed pattern of local multipartite measurements with \(W=1\); for the honeycomb lattice, Perseguers et al.\ report the threshold \(2\sin^2\theta_{\mathrm{th}}=0.609(0)\)~\cite{perseguers2010multipartite}.
CEP and the Perseguers GHZ construction therefore compare different processing strategies under the same \(w=1\), \(W=1\) resource-layer supply.

In two-copy CEP, the two elementary pairs at each honeycomb link location undergo joint singlet conversion with optimal success probability \(\min\{1,2(1-\cos^4\theta)\}\)~\cite{acin2007entanglement}.
The Acín two-copy construction begins with two elementary pairs per honeycomb link location and transforms the bulk honeycomb network into an effective triangular lattice before applying CEP~\cite{acin2007entanglement}.
Under the common \(w=1\) normalization, both two-copy CEP and the Acín two-copy construction use \(W=2\), with elementary-pair thresholds \(0.35848\) and \(0.34730\), respectively.
Thus this comparison changes the processing strategy while keeping the elementary-pair supply and aggregation window fixed.

\paragraph{Fixed-window entanglement transport.}
The framework converts the fixed-window connectivity transitions for CEP, the Perseguers GHZ construction, two-copy CEP, and the Acín two-copy construction into distance-weighted entanglement transport scaling.
For each of these constructions, the three honeycomb link locations incident to either requested endpoint supply at most \(3W\) elementary pairs across the corresponding endpoint cut.
Since \(F_\star=1\), the Schmidt-rank argument in Appendix~\ref{app:percolation-comparison} gives \(B_1^{(W)}\le 3W\).
For fixed \(\theta\) strictly below the relevant threshold, connectivity in the corresponding percolation problem decays exponentially with separation~\cite{aizenman1987sharpness}.
This gives \(O(N^{-1})\) entanglement transport per activation slot, while admissible bounded-distance sessions give the matching \(\Omega(N^{-1})\) order.
For fixed \(\theta\) strictly above the threshold, supercritical percolation on this resource-graph family gives an \(N\)-independent positive connection probability for a positive fraction of macroscopically separated admissible endpoint pairs~\cite{kesten1982percolation,grimmett1999percolation}, yielding an \(\Omega(\sqrt N)\) lower bound, while the endpoint bound and the \(O(\sqrt N)\) graph diameter give the matching upper bound.
Hence CEP, the Perseguers GHZ construction, two-copy CEP, and the Acín two-copy construction have entanglement transport \(\Theta(N^{-1})\) below threshold and \(\Theta(\sqrt N)\) above threshold under their respective fixed aggregation windows.
Appendix~\ref{app:percolation-comparison} gives the derivation and the endpoint qualifications for the Perseguers GHZ and Acín two-copy constructions.

\paragraph{Concurrence percolation theory.}
Concurrence percolation theory (ConPT) evaluates connectivity using concurrence instead of singlet-conversion probability, with one elementary pair per honeycomb link location~\cite{meng2021concurrence}.
Its reported honeycomb threshold \(\theta_{\mathrm{th}}/(\pi/4)=0.51(8)\) corresponds to the elementary-pair singlet-conversion probability shown in Table~\ref{tab:percolation-comparison}.
However, the higher-order star--mesh connectivity rules used to evaluate ConPT are approximations that may not correspond to LOCC-realizable transformations~\cite{meng2021concurrence}.
Accordingly, Table~\ref{tab:percolation-comparison} does not assign operational entanglement transport scaling orders to ConPT on the honeycomb lattice.

\paragraph{GCP direct all-shortest-path realization.}
Nath and Roy's general concurrence percolation (GCP) assigns effective entangled connections to node pairs using the concurrence series--parallel rules over all shortest paths and reports the honeycomb threshold \(\theta_{\mathrm{th}}/(\pi/4)=0.407(5)\), corresponding to elementary-pair singlet-conversion probability \(2\sin^2\theta_{\mathrm{th}}=0.197(5)\)~\cite{nath2026general}.
Under the comparison assumptions, exact simultaneous preparation of one effective pure two-qubit state for every unordered node pair requires \(W=\Omega(N^{3/2})\).
Table~\ref{tab:percolation-comparison} evaluates the direct realization that supplies distinct elementary-pair copies to every prescribed path use and retains only the effective states for final processing.
Its minimum aggregation window is \(W(N)=\exp[\Theta(\sqrt{N})]\), and independent singlet conversion followed by maximum pure-Bell-pair routing achieves entanglement transport \(\exp[-\Theta(\sqrt{N})]\) per activation slot.
Appendix~\ref{app:gcp-direct-realization} gives the preparation bounds and the analysis of this realization.

\section{Chain network benchmark}
\label{sec:chain-benchmark}

We consider the homogeneous \(N\)-node chain with spacing \(a_N\), random exclusive pairing and fixed link multiplicity \(w\).
Orient each session so that \(s_i<t_i\).
Link opportunity outcomes are independent across link opportunities and activation slots, with common success probability \(p\in(0,1)\).
Each realized link supplies an ideal Bell pair.
Resource-layer processing and storage are ideal, and strategies may use all heralded realized link information.
Both finite-window and asymptotic aggregation regimes use the same fidelity threshold \(F_\star=0.95>F_{\min}\).
As \(N\to\infty\), a limit stated uniformly over even \(2\le M\le N\) holds with the maximum deviation over these \(M\) tending to zero, and an order stated uniformly holds with constants independent of \(M\).

\subsection{Fixed-window and asymptotic aggregation}
\label{sec:chain-resource}

Table~\ref{tab:chain_scaling} gives the scaling laws of the chain network in finite-window and asymptotic aggregation regimes under expansion and densification.
Session normalization divides each capacity by \(|\mathcal D|=M/2\), as in Eq.~\eqref{eq:session-normalized-capacity}.
For fixed \(W\), delivery of a fidelity-qualified pair requires every link location along the session's path to supply an elementary pair within the window.
The probability of this event decays exponentially with hop distance, so short sessions determine the leading fixed-window transport order, whereas asymptotic aggregation permits delivery over \(\Theta(N)\) hops. 
Appendix~\ref{app:chain-benchmark} gives the derivation and the proofs of the theorems below.

\begin{theorem}[Fixed-window capacity]
\label{thm:chain-single-slot}
For every finite \(W\ge1\),
\begin{equation}
C_{\mathrm{ET}}^{(W)}(N,M;1)
\le
C_{\mathrm{ET}}^{(W)}(N,M;0.95)
\le
\kappa(0.95)C_{\mathrm{ET}}^{(W)}(N,M;1),
\label{eq:chain-window-fidelity-comparison}
\end{equation}
where \(\kappa(0.95)\simeq1.5764\) is the fidelity calibration factor in Eq.~\eqref{eq:fidelity-calibration-factor}.
For fixed \(w\), \(p\in(0,1)\) and \(W\) independent of \(N\), as \(N\to\infty\), uniformly over even \(2\le M\le N\),
\begin{equation}
C_{\mathrm{ET}}^{(W)}(N,M;0.95)
=
\Theta\!\left(\frac{Ma_N}{N}\right).
\label{eq:chain-fixed-window-scaling}
\end{equation}
\end{theorem}

The lower bound in Eq.~\eqref{eq:chain-window-fidelity-comparison} follows because every perfect Bell pair satisfies the threshold \(F_\star=0.95\).
The upper bound is proved in Appendix~\ref{app:chain-benchmark}.
Under the fixed-parameter and uniformity conditions of Theorem~\ref{thm:chain-single-slot},
\begin{equation}
\frac{N}{Ma_N}C_{\mathrm{ET}}^{(W)}(N,M;1)
\longrightarrow
\frac{\Gamma_{wW,p}}{W},
\qquad N\to\infty,
\label{eq:chain-window-perfect-reference}
\end{equation}
uniformly over even \(2\le M\le N\), where \(\Gamma_{wW,p}\in(0,\infty)\) is defined in Appendix~\ref{app:chain-benchmark}.
For general \(w\) and fixed finite \(W\), the fidelity-calibrated bounds in Eq.~\eqref{eq:chain-window-fidelity-comparison} establish the same \(\Theta(Ma_N/N)\) order at \(F_\star=0.95\) but do not determine the corresponding normalized large-\(N\) limit.

\begin{proposition}[Exact single-slot identity at the benchmark point]
\label{prop:fq-single-slot}
For \(w=1\), \(W=1\) and \(F_\star=0.95\), with all heralded realized link information,
\[
C_{\mathrm{ET}}^{(1)}(N,M;0.95)=C_{\mathrm{ET}}^{(1)}(N,M;1).
\]
\end{proposition}

\begin{theorem}[Asymptotic aggregation capacity]
\label{thm:chain-asymptotic}
For fixed \(w\), \(p\in(0,1)\) and \(F_\star=0.95\), as \(N\to\infty\), uniformly over even \(2\le M\le N\),
\[
C_{\mathrm{ET}}^{(\infty)}(N,M;0.95)
=
\Theta(wp\,a_NN).
\]
\end{theorem}

For fixed \(w\), \(p\in(0,1)\) and \(W\) independent of \(N\), as \(N\to\infty\), uniformly over even \(2\le M\le N\), the two theorems therefore give
\begin{equation}
\frac{C_{\mathrm{ET}}^{(\infty)}(N,M;0.95)}{C_{\mathrm{ET}}^{(W)}(N,M;0.95)}
=
\Theta\!\left(\frac{N^2}{M}\right),
\label{eq:chain-aggregation-separation}
\end{equation}

In expansion, session normalization makes the aggregation contrast explicit: fixed-window capacity per requested session scales as \(\Theta(N^{-1})\), whereas asymptotic-aggregation capacity per requested session scales as \(\Theta(N/M)\).
Thus, for a fixed number of active users, fixed-window capacity per session vanishes as the network grows, while asymptotic-aggregation capacity per session increases linearly with network size.

\begin{table}[t]
\centering
\caption{Entanglement transport capacity of the homogeneous chain with fixed \(w\) and \(p\in(0,1)\), independent ideal Bell-pair supply and ideal resource processing and storage.
Both capacities use all heralded realized link information and \(F_\star=0.95\).
The orders hold as \(N\to\infty\) after averaging over random exclusive pairing, uniformly over even \(2\le M\le N\).
The fixed-window column uses \(W\) independent of \(N\), and \(a>0\) and \(L>0\) are fixed in their respective growth regimes.}
\label{tab:chain_scaling}
\begin{tabular}{lcc}
\toprule
Regime & Fixed-window \(C_{\mathrm{ET}}^{(W)}\) & Asymptotic \(C_{\mathrm{ET}}^{(\infty)}\) \\
\midrule
Expansion: \(a_N=a\) & \(\Theta(M/N)\) & \(\Theta(N)\) \\
Densification: \(a_N=L/(N-1)\) & \(\Theta(M/N^2)\) & \(\Theta(1)\) \\
\bottomrule
\end{tabular}
\end{table}

\subsection{Scaling under physical implementations}
\label{sec:chain-physical}

A session is \textit{isolated} when it is served alone, without sharing its link locations with any other session.
For a fixed fidelity threshold \(F_\star\) and an isolated session \(i\) at hop distance \(h\), when the long-run limit exists, define the fidelity-qualified Bell-pair delivery rate \(b_h\) and the physical hop-profile coefficient \(\Lambda\) by
\begin{equation}
b_h
:=
\lim_{T\to\infty}
\frac{\mathbb E[B_i(T)\mid h_i=h]}{T},
\qquad
\Lambda
:=
\sum_{h\ge1}h b_h.
\label{eq:physical-hop-profile}
\end{equation}

For the given demand, an isolated session \(i\) has entanglement-transport rate \(a_N h_i b_{h_i}\).
Before observing physical link outcomes, select a pairwise link-disjoint subset that maximizes the sum of these rates.
This defines the pre-heralding construction used below.

\begin{theorem}[Physical hop-profile scaling law]
\label{thm:hop-profile}
Fix the expansion regime and all physical parameters, and let \(\{b_h\}_{h\ge1}\) be an \(N\)-independent hop profile satisfying \(0<\Lambda<\infty\).
Under this construction, assume that the selected sessions operate concurrently with independent link-opportunity outcomes, sufficient local resources, and no reduction of their isolated-session rates.
Then, uniformly over even \(2\le M\le N\),
\[
\frac{N}{Ma_N}\underline S_{\mathrm{ET}}(N,M;F_\star)
\longrightarrow
\Lambda,
\qquad N\to\infty.
\]
\end{theorem}
Appendix~\ref{app:physical-multiuser} gives the derivation.

The physical benchmarks use expansion with \(a_N=a=1\,\mathrm{km}\), \(w=1\), \(F_\star=0.95\), common link opportunity probability \(p=0.05\exp[-a/(22\,\mathrm{km})]\), and reference interval \(1/r=45\,\mu\mathrm{s}\).
The matched noisy physical implementations use the initialization, memory-dephasing, effective swapping-error, and measurement-error maps and the operation-time budgets described in Appendix~\ref{app:matched-noisy-hardware}, drawing on Kinos et al.~\cite{kinos2025complete}, while the retention, timing, and provisioned parallel operation at neighboring link locations are specified by this benchmark.
As a deliberate simplification, all physical implementations neglect path-wide signaling delays: generation stops, and purification or swapping starts, once every link location on the session path has supplied the required elementary pairs, and the swapping outcomes determining the endpoint Pauli corrections are available at swap completion.
Herald propagation within each reference interval and, for the one-round DEJMPS implementation, neighboring purification-outcome exchange are included in the timing.

Hardware noise lowers the fidelity of the delivered pair: initialization errors and memory dephasing degrade each elementary pair, and swapping and measurement errors accumulate over the \(h-1\) swaps.
For the noisy physical implementation, these effects define an \(h\)-dependent \textit{waiting budget}: a delivered pair is fidelity-qualified if and only if the total waiting time of its retained elementary pairs, counted in activation slots from their heralds until every link location on the session path has supplied an elementary pair, does not exceed this budget.
At the benchmark point, the waiting budget decreases from \(495.2\) activation slots at \(h=2\) to \(60.9\) at \(h=9\) and is negative for \(h\ge10\), so \(b_h=0\) for \(h\ge10\) and \(\Lambda\) is finite (Appendix~\ref{app:matched-noisy-hardware}).
Up to the hop ceiling, the waiting budget grows linearly with the memory coherence time, and its decrease with \(h\) makes \(b_h\) fall sharply as \(h\) approaches the hop ceiling.
The one-round DEJMPS physical implementation raises the elementary-pair fidelity at the cost of additional elementary pairs and time and extends the hop ceiling to \(12\).

At this benchmark point, Proposition~\ref{prop:fq-single-slot} allows the single-slot reference ratio in Eq.~\eqref{eq:single-slot-reference-ratio} to be evaluated using the perfect-output capacity.
For \(0<\Lambda<\infty\), under the same asymptotic conditions as Theorem~\ref{thm:hop-profile},
\begin{equation}
\frac{\underline S_{\mathrm{ET}}}{rC_{\mathrm{ET}}^{(1)}}
\longrightarrow
\frac{\Lambda}{r\Gamma_{1,p}},
\qquad N\to\infty,
\qquad
\Gamma_{1,p} = \frac{p}{(1-p)^2}.
\label{eq:physical-single-slot-ratio}
\end{equation}
The numerical ratios in Table~\ref{tab:chain-physical} are Monte Carlo estimates of this large-\(N\) limit, not exact finite-\(N\) service ratios. 
\(\widehat\Lambda\) denotes the Monte Carlo estimate of \(\Lambda\).

\begin{table}[t]
\centering
\caption{Physical chain benchmarks in expansion with \(a_N=a=1\,\mathrm{km}\), \(w=1\), \(F_\star=0.95\), homogeneous link opportunity probability \(p=0.05\exp[-a/(22\,\mathrm{km})]\), and reference interval \(1/r=45\,\mu\mathrm{s}\). 
The noisy physical implementation retains one elementary pair per link location and uses no purification. 
The one-round DEJMPS physical implementation retains three elementary pairs per link location. 
Ideal retention assumes ideal preparation, memories, and operations, zero swapping duration, and at most one retained elementary pair per link location.
All physical implementations neglect path-wide signaling delays.
The hop ceilings are established by deterministic fidelity bounds. 
For the two noisy rows, \((N/(Ma_N))\underline S_{\mathrm{ET}}\to\Lambda\) as \(N\to\infty\) by Theorem~\ref{thm:hop-profile}.
Parentheses on \(\widehat\Lambda\) values give one renewal-reward standard error, and brackets on the achieved single-slot reference ratios give nominal asymptotic-normal \(95\%\) intervals. 
For ideal retention, \(b_h=\Theta(rp/\log h)\) is for \(h\to\infty\) at fixed \(p\in(0,1)\), while the achieved-service and achieved-single-slot-reference-ratio orders are for \(N\to\infty\), uniformly over even \(2\le M\le N\), with the stated physical parameters fixed.}
\label{tab:chain-physical}
\footnotesize
{\setlength{\tabcolsep}{4pt}
\renewcommand{\arraystretch}{1.2}
\begin{tabularx}{\textwidth}{@{}Xm{0.36\linewidth}m{0.16\linewidth}m{0.16\linewidth}@{}}
\hline
Physical implementation
& Hop profile
& \(\underline S_{\mathrm{ET}}\)
& \(\underline S_{\mathrm{ET}}/(rC_{\mathrm{ET}}^{(1)})\)
\\
\hline
Noisy hardware
& \(0<\Lambda<\infty\); deterministic hop ceiling \(9\); \(\widehat\Lambda=15144.5(3.2)\,\mathrm{s}^{-1}\)
& \(\Theta(Ma_N/N)\)
& \( \underset{[12.9282,12.9388]}{12.9335}\)
\\
One-round DEJMPS
& \(0<\Lambda<\infty\); deterministic hop ceiling \(12\); \(\widehat\Lambda=17992.9(3.4)\,\mathrm{s}^{-1}\)
& \(\Theta(Ma_N/N)\)
& \(\underset{[15.3603,15.3717]}{15.3660}\)
\\
Ideal retention
& \(b_h=\Theta(rp/\log h)\); \(\Lambda=\infty\)
& \(\Theta(rpa_NN/\log N)\)
& \(\Theta(N^2/(M\log N))\)
\\
\hline
\end{tabularx}}
\end{table}

Theorem~\ref{thm:hop-profile} therefore gives constructive \(\Theta(M/N)\) expansion scaling for both matched noisy physical implementations, with different coefficients.
Appendix~\ref{app:physical-numerics} gives the hop profiles and numerical uncertainty analysis.

The ideal-retention physical implementation assumes ideal preparation, memories, and operations, zero swapping duration, and at most one retained elementary pair per link location per delivery cycle.
For fixed \(p\in(0,1)\), this gives
\[
b_h
=
\Theta\!\left(\frac{rp}{\log h}\right),
\qquad
\Lambda=\infty,
\]
as \(h\to\infty\), so Theorem~\ref{thm:hop-profile} does not apply.
Under the pre-heralding maximum-weight pairwise link-disjoint construction, as \(N\to\infty\) with \(a_N=a\), \(p\), and \(r\) fixed, uniformly over even \(2\le M\le N\),
\begin{equation}
\underline S_{\mathrm{ET}}
=
\Theta\!\left(\frac{rpa_NN}{\log N}\right).
\label{eq:ideal-retention-service-main}
\end{equation}
The link locations wait until the other link locations are ready while retaining at most one elementary pair, producing the logarithmic time penalty.
Equation~\eqref{eq:ideal-retention-service-main} relies on neglecting path-wide signaling delays: signaling across an \(h\)-hop session path takes time proportional to \(h\), which exceeds the logarithmic waiting time for sufficiently large \(h\).
At reference slot rate \(r\), asymptotic aggregation satisfies \(rC_{\mathrm{ET}}^{(\infty)}(N,M;0.95)=\Theta(rpa_NN)\) for \(w=1\), so the ideal-retention physical implementation has a \(\Theta(\log N)\) gap to asymptotic aggregation.
Appendix~\ref{app:ideal-retention} gives the derivation.

To compare performance per requested session, divide the achieved service by \(|\mathcal D|=M/2\), analogously to the resource-layer session normalization in Eq.~\eqref{eq:session-normalized-capacity}.
For the noisy implementations with \(0<\Lambda<\infty\),
\[
\frac{N}{a_N}\,\frac{\underline S_{\mathrm{ET}}}{|\mathcal D|}
\longrightarrow
2\Lambda,
\qquad N\to\infty,
\]
whereas ideal retention gives
\[
\frac{\underline S_{\mathrm{ET}}}{|\mathcal D|}
=
\Theta\!\left(
\frac{rpa_NN}{M\log N}
\right).
\]
Thus, in fixed-spacing expansion with a fixed number of active users, service per requested session decreases as the network grows for both noisy implementations, whereas it increases under ideal retention.
If the number of active users also grows with the network, the ideal-retention per-session scaling depends on that growth through the factor \(N/(M\log N)\).

\section{Discussion}
\label{sec:discussion}

Entanglement transport capacity per activation slot and service per second provide a common basis for studying quantum-network scaling through the same fidelity-qualified, distance-weighted delivery objective.
The resource layer captures topology, demand, supply, and aggregation constraints, while the physical layer incorporates implementation constraints, imperfections, and wall-clock timing.
Their matched comparison gives a service ceiling under the stated resource-accounting conditions.
The percolation classification compares the examined entanglement and concurrence percolation constructions under common resource accounting.
It shows why thresholds alone do not determine transport performance, which also depends on resource supply, aggregation, and admissible processing.

In the chain network benchmark, fixed finite-window entanglement transport is dominated by short sessions because the probability of supplying a long path decays exponentially with hop distance.
Asymptotic aggregation supports long sessions by accumulating elementary pairs across arbitrarily many slots under ideal storage.
Both noisy physical implementations retain hop ceilings independent of network size, so purification changes the coefficient but not the achieved scaling.
Ideal retention permits long sessions, but waiting for every link location to supply an elementary pair while retaining at most one pair per link location produces a logarithmic time penalty.

Extending the framework to more general network families is a promising direction for future research.
For the percolation constructions in Section~\ref{sec:classification}, future work includes deriving upper bounds on entanglement transport capacity under asymptotic aggregation, evaluating achieved service under physical implementations, and extending the comparison to other lattices, such as the square lattice.
Unlike the fixed-window chain benchmark, supercritical lattice percolation can retain long-distance connectivity through alternative paths.
A central question is whether a given physical implementation on the lattice imposes a hop ceiling independent of network size and, if so, how this restriction changes its service scaling.

Satellite links add long-range connections that bypass terrestrial hops~\cite{yin2017satellite}, motivating a study of their effect on entanglement transport scaling under specified resource growth.
Periodically varying link-opportunity probabilities provide a model of recurring orbital passes beyond the slotwise iid baseline.
Physical-service analysis must coordinate satellite and terrestrial link generation while accounting consistently for changing optical losses, heralding delays, memory aging, and final fidelity qualification.

\paragraph{Acknowledgment.}
We thank Jan Østergaard and Shashi Raj Pandey for a careful reading of the manuscript and detailed comments.
This work is supported by Danish National Research Foundation (DNRF) through the Center of Excellence CLASSIQUE, grant no. DNRF187.

\paragraph{Data and code availability.}
The code, numerical data, and verification materials supporting the terrestrial physical-layer benchmark are archived on Zenodo at \href{https://doi.org/10.5281/zenodo.23031358}{doi:10.5281/zenodo.23031358}.
The archive corresponds to release version \texttt{1.1.0} and calculation configuration \texttt{20260905.1}.

\paragraph{AI-assisted research and manuscript preparation.}
Generative-AI tools were used during the development of this work.
OpenAI ChatGPT (GPT-5.6 and GPT-6 Astra) and Anthropic Claude (Opus 5 and Opus 5.5) assisted with scientific reasoning and derivation checking, literature search and source verification, manuscript organization and drafting, preparation of \LaTeX{} and TikZ source, and the development, execution, and verification of the numerical calculations underlying the reproducibility package.
The authors directed these uses, checked the analytical arguments and source-dependent claims against independent derivations and primary literature, reviewed and revised AI-assisted manuscript text, and take responsibility for the scientific claims and final manuscript.
The role and verification of AI-assisted numerical and reproducibility-package work are described separately in the package.

\medskip
\printbibliography
\clearpage

\appendix
\section*{Appendix}

\section{Fidelity calibration for cut constraints}
\label{app:fidelity-calibration}

Fix a resource graph, a demand \(\mathcal D\), an aggregation window of finite length \(W\ge1\), and a cut separating the network nodes into two sets.
Let \(E_{\mathrm{sq}}\) denote squashed entanglement across the chosen cut.
Take \(F_\star\in(F_{\min},1)\) and use the fidelity calibration factor \(\kappa(F_\star)\) defined in Eq.~\eqref{eq:fidelity-calibration-factor}.

Condition on the heralded realized link information and the classical delivery record identifying the output systems and sessions, with unobserved noise averaged as in the framework's qualification rule.
Let \(K\) be the number of fidelity-qualified Bell pairs crossing the cut, let \(\rho_{\mathrm{out}}\) be their joint conditional state after all other systems are discarded, and let \(\rho_j\) be its two-qubit marginal for the \(j\)th counted Bell pair, with \(F_j=\bra{\Phi^+}\rho_j\ket{\Phi^+}\ge F_\star\).
Isotropic twirling preserves \(F_j\) and produces a Bell-diagonal state with eigenvalues \(F_j,(1-F_j)/3,(1-F_j)/3,(1-F_j)/3\).
Since squashed entanglement is LOCC monotone and upper-bounds distillable entanglement, the one-way hashing bound gives~\cite{bennett1996mixed,christandl2004squashed}
\begin{equation}
E_{\mathrm{sq}}(\rho_j)
\ge 1+F_j\log_2F_j+(1-F_j)\log_2\!\left(\frac{1-F_j}{3}\right)
\ge\frac{1}{\kappa(F_\star)}.
\label{eq:app-fidelity-single-pair}
\end{equation}
The middle expression has derivative \(\log_2[3F_j/(1-F_j)]>0\) for \(1/4<F_j<1\), and its value at \(F_j=1\) is understood by continuity.
Superadditivity of squashed entanglement applies to the marginals of the common conditional joint state, including when the counted Bell pairs are correlated, and yields~\cite{christandl2004squashed}
\begin{equation}
E_{\mathrm{sq}}(\rho_{\mathrm{out}})
\ge\sum_{j=1}^{K}E_{\mathrm{sq}}(\rho_j)
\ge\frac{K}{\kappa(F_\star)}.
\label{eq:app-fidelity-correlated-output}
\end{equation}

For fixed \(\mathcal D\) and heralded realized link information, suppose that the strategy can be simulated by LOCC on a joint supplied resource state \(\rho_{\mathrm{in}}\), with each side of the cut treated as one party.
Under the framework's independent supply model, \(\rho_{\mathrm{in}}\) can be taken to contain all elementary pairs supplied within the initially empty aggregation window, with unobserved noise averaged.
LOCC monotonicity on average therefore gives~\cite{christandl2004squashed}
\begin{equation}
\mathbb E[K]
\le\kappa(F_\star)\mathbb E\!\left[E_{\mathrm{sq}}(\rho_{\mathrm{out}})\right]
\le\kappa(F_\star)E_{\mathrm{sq}}(\rho_{\mathrm{in}}),
\label{eq:app-fidelity-general-cut}
\end{equation}
where these expectations average over delivery records and strategy randomization at fixed \(\mathcal D\) and heralded realized link information.
The inequality can then be averaged over the realized links and demand, and divided by \(W\) to obtain a bound per activation slot.
When every supplied elementary pair is an ideal Bell pair, \(E_{\mathrm{sq}}(\rho_{\mathrm{in}})\) equals the number of supplied elementary pairs crossing the cut~\cite{christandl2004squashed}.
For other declared elementary-pair states, the converse requires a valid upper bound on the squashed entanglement of the joint supplied state across the cut.
This calibration applies to arbitrary resource graphs; identifying the resulting cut constraints with an achievable capacity requires a separate argument.

\section{Percolation-family threshold conversion and resource accounting}
\label{app:percolation-comparison}

\subsection{Threshold conversion on the honeycomb lattice}

For the elementary state \(\ket{\psi(\theta)}=\cos\theta\ket{00}+\sin\theta\ket{11}\), the optimal single-copy singlet-conversion probability is \(2\sin^2\theta\)~\cite{acin2007entanglement}.

For one-copy CEP, the exact honeycomb bond-percolation threshold gives
\begin{equation}
2\sin^2\theta_{\mathrm{th}}^{\mathrm{CEP}} = 1-2\sin\!\left(\frac{\pi}{18}\right) = 0.6527036447\ldots.
\label{eq:cep-honeycomb-threshold}
\end{equation}

For the Perseguers GHZ construction, the same quantity parametrizes both the elementary-pair entanglement and the site-percolation probability of the transformed construction, and the source reports
\begin{equation}
2\sin^2\theta_{\mathrm{th}}^{\mathrm{GHZ}} = 0.609(0)
\label{eq:perseguers-ghz-threshold}
\end{equation}
for the honeycomb lattice~\cite{perseguers2010multipartite}.

For two elementary pairs at one honeycomb link location, the largest eigenvalue of either reduced state is \(\cos^4\theta\), so in the threshold range the optimal singlet-conversion probability of the two-copy state is \(2(1-\cos^4\theta)\)~\cite{acin2007entanglement}.
Equating this two-copy conversion probability to the honeycomb bond threshold gives
\begin{equation}
2\sin^2\theta_{\mathrm{th}}^{\mathrm{CEP},2}
=
2-\sqrt{2+4\sin\!\left(\frac{\pi}{18}\right)}
=
0.3584785379\ldots.
\label{eq:two-copy-cep-threshold}
\end{equation}

In the Acín two-copy construction, the topology-changing measurements generate an effective triangular lattice whose connections have singlet-conversion probability \(2\sin^2\theta\)~\cite{acin2007entanglement}.
The resulting threshold is therefore
\begin{equation}
2\sin^2\theta_{\mathrm{th}}^{\mathrm{Acin}}
=
2\sin\!\left(\frac{\pi}{18}\right)
=
0.3472963553\ldots.
\label{eq:acin-honeycomb-threshold}
\end{equation}

Meng et al.\ report the honeycomb ConPT estimate \(\theta_{\mathrm{th}}/(\pi/4)=0.51(8)\)~\cite{meng2021concurrence}.
Under the common elementary-pair parametrization this gives
\begin{equation}
2\sin^2\theta_{\mathrm{th}}^{\mathrm{ConPT}}
\simeq
0.30,
\label{eq:conpt-honeycomb-threshold}
\end{equation}
with the source interval \(0.51\pm0.08\) mapping approximately to \([0.22,0.40]\), reported in Table~\ref{tab:percolation-comparison} as \(0.30^{+0.10}_{-0.08}\).

Nath and Roy report the honeycomb GCP estimate \(\theta_{\mathrm{th}}/(\pi/4)=0.407(5)\)~\cite{nath2026general}.
The corresponding elementary-pair singlet-conversion probability is
\begin{equation}
2\sin^2\theta_{\mathrm{th}}^{\mathrm{GCP}}
\simeq
0.197,
\label{eq:gcp-honeycomb-threshold}
\end{equation}
and the reported angle interval maps approximately to \([0.193,0.202]\), reported in Table~\ref{tab:percolation-comparison} as \(0.197(5)\).

\subsection{Fixed-window single-session entanglement transport}

Throughout this subsection, asymptotics are taken as \(N\to\infty\) along the expanding honeycomb resource-graph family defined in Section~\ref{sec:classification}, with \(\theta\) fixed strictly away from the relevant threshold and \(W\) fixed independently of \(N\).
For a uniformly sampled session, the honeycomb hop distance \(h_1\) and physical separation \(d(s_1,t_1)\) are comparable up to \(N\)-independent constants, the network diameter is \(\Theta(\sqrt N)\), and the number of unordered endpoint pairs with \(h_1=h\) is \(O(Nh)\).
A positive fraction of all unordered endpoint pairs have \(h_1=\Theta(\sqrt N)\) and \(d(s_1,t_1)=\Theta(\sqrt N)\).

The fixed-window constructions considered here are CEP, the Perseguers GHZ construction, two-copy CEP, and the Acín two-copy construction.
For the Perseguers GHZ construction, the local measurements prescribed in Ref.~\cite{perseguers2010multipartite} leave a fixed nonzero fraction of the honeycomb nodes available as long-distance endpoints. 
For uniformly random demand, we apply the construction when both requested endpoints belong to this retained set and deliver no pair otherwise.
For the Acín two-copy construction, half of the honeycomb nodes undergo the topology-changing measurement and the other half form the effective triangular lattice~\cite{acin2007entanglement}. 
For uniformly random demand, we apply the construction when both requested endpoints belong to this retained set and deliver no pair otherwise.
Because each retained set contains an \(N\)-independent fraction of the nodes, these endpoint restrictions change only constant factors and not the percolation thresholds or the resulting \(N\)-scaling orders.

For each construction, the three honeycomb link locations incident to either requested endpoint supply at most \(3W\) elementary pairs across the corresponding endpoint cut, so the initial Schmidt rank across that cut is at most \(2^{3W}\).
Decompose each conditional local operation into single-Kraus branches.
Conditioned on a complete sequence of such branches, the global state remains pure, and local operations cannot increase its Schmidt rank across the endpoint cut.
Since \(F_\star=1\), every pair counted by \(B_1^{(W)}\) is an ideal Bell pair and therefore factors from the remaining systems on the branch.
A branch delivering \(B_1^{(W)}\) counted pairs consequently has Schmidt rank at least \(2^{B_1^{(W)}}\), so \(B_1^{(W)}\le3W\).

Below the corresponding percolation threshold, standard subcritical percolation gives
\[
\Pr\!\left(B_1^{(W)}\ge1\mid h_1=h\right)=O(e^{-ch})
\]
for some \(N\)-independent \(c>0\)~\cite{aizenman1987sharpness}.
Together with \(B_1^{(W)}\le3W\), this gives
\[
\mathbb E\!\left[B_1^{(W)}\mid h_1=h\right]=O\!\left(We^{-ch}\right).
\]
Because \(O(Nh)\) unordered sessions have hop distance \(h\) among \(\Theta(N^2)\) possible sessions, their contribution to entanglement transport is bounded by
\begin{equation}
\frac{1}{W}\mathbb E\!\left[d(s_1,t_1)B_1^{(W)}\right]
=
O\!\left(
\frac{1}{WN}\sum_{h\ge1}h^2We^{-ch}
\right)
=
O\!\left(
\frac{1}{N}\sum_{h\ge1}h^2e^{-ch}
\right)
=
O(N^{-1}).
\label{eq:percolation-subcritical-upper}
\end{equation}
For each construction, there are \(\Theta(N)\) admissible bounded-distance sessions for which the corresponding finite local construction succeeds with an \(N\)-independent positive probability at fixed \(\theta>0\).
These sessions give the matching \(\Omega(N^{-1})\) lower bound.

Above the corresponding percolation threshold, the relevant bond- or site-percolation construction has a strictly positive long-distance connection probability for admissible endpoints~\cite{acin2007entanglement,perseguers2010multipartite}.
For the resource-graph family used here, applying the supercritical finite-size property of percolation on the corresponding periodic two-dimensional lattice, namely the honeycomb lattice, the effective triangular lattice, or the transformed lattice of Ref.~\cite{perseguers2010multipartite}, gives an \(N\)-independent positive connection probability for a positive fraction of macroscopically separated admissible endpoint pairs~\cite{kesten1982percolation,grimmett1999percolation}.
A uniformly random session belongs to this class with probability \(\Theta(1)\), and a successful connection delivers at least one ideal Bell pair.
Since these pairs have \(d(s_1,t_1)=\Theta(\sqrt N)\) and \(W\) is fixed, they give an \(\Omega(\sqrt N)\) entanglement transport lower bound.
The endpoint-cut bound \(B_1^{(W)}=O(1)\) and the \(O(\sqrt N)\) graph diameter give the matching upper bound.
Hence the four constructions achieve the following orders:
\begin{equation}
\frac{1}{W}\mathbb E\!\left[d(s_1,t_1)B_1^{(W)}\right]
=
\begin{cases}
\Theta(N^{-1}), & \theta<\theta_{\mathrm{th}},\\[1mm]
\Theta(\sqrt N), & \theta>\theta_{\mathrm{th}},
\end{cases}
\label{eq:fixed-window-percolation-et}
\end{equation}
with the threshold and fixed \(W\) corresponding to the construction under consideration.

No statement here concerns the critical points themselves.
At the critical point, connectivity need not decay exponentially, so its critical decay can interact with the growing distance weight and vanishing long-range connection probability alone does not determine the entanglement transport scaling.

\subsection{Aggregation and entanglement transport for a direct GCP realization}
\label{app:gcp-direct-realization}

We analyze an explicit operational realization of the GCP calculation underlying Table~\ref{tab:percolation-comparison}.
For each non-neighboring node pair, Nath and Roy combine the contributions of all shortest paths using the concurrence series rule along each path and the parallel rule across paths before singlet conversion, while neighboring pairs use their direct elementary state~\cite{nath2026general}.
In the direct realization considered here, we assign a distinct elementary-pair copy to every prescribed use of a link location.
We apply the parallel rule~\cite[Eq.~(5)]{nath2026general} to the tensor product of the resulting path states.
After all pairwise preparations, we retain one effective pure two-qubit state for each unordered node pair and discard all other quantum systems before final processing.
The retained states may then undergo arbitrary final LOCC, with independent singlet conversion followed by maximum pure-Bell-pair routing as one admissible strategy.
We inherit \(F_\star=1\), \(w=1\), \(p_e=1\), \(M=2\), fixed \(0<\theta<\pi/4\), empty initial windows, ideal resource-layer storage and LOCC, and fixed nearest-neighbor physical spacing from Section~\ref{sec:classification}.

For distinct nodes \(u,v\), let \(\mathcal P_{uv}\) denote the set of shortest paths between them, and define the number of prescribed elementary-pair uses at link location \(e\) by
\begin{equation}
\ell_e
:=
\sum_{\{u,v\}\subseteq V}
\sum_{P\in\mathcal P_{uv}}
\mathbf 1\{e\in P\}.
\label{eq:gcp-link-load}
\end{equation}

\begin{proposition}[Direct GCP realization]
\label{prop:gcp-direct-realization}
As \(N\to\infty\) along the same expanding honeycomb resource-graph family, with \(\theta\) fixed, the minimum aggregation window \(W(N)\) that supplies all elementary-pair copies prescribed above satisfies
\begin{equation}
W(N)
=
\max_{e\in E}\ell_e
=
\exp[\Theta(\sqrt N)].
\label{eq:gcp-window-scaling}
\end{equation}
If the final processing is optimized over arbitrary LOCC acting only on the retained effective states, then
\begin{equation}
\sup_{\mathrm{final\ LOCC}}
\frac{1}{W(N)}
\mathbb E\!\left[
d(s_1,t_1)B_1^{(W(N))}
\right]
=
\exp[-\Theta(\sqrt N)].
\label{eq:gcp-final-locc-scaling}
\end{equation}
Independent singlet conversion followed by maximum pure-Bell-pair routing attains the same stretched-exponential order.
The expectation in Eq.~\eqref{eq:gcp-final-locc-scaling} is over the uniformly sampled session and all processing outcomes.
\end{proposition}

\paragraph{Exact aggregation accounting.}
A window of \(W\) activation slots supplies exactly \(W\) elementary-pair copies at each link location because \(w=1\) and \(p_e=1\).
The prescribed path computations require exactly \(\ell_e\) distinct copies at location \(e\), so \(W(N)\ge\max_e\ell_e\).
Conversely, after \(\max_e\ell_e\) slots, every link location has supplied enough copies to assign one to every prescribed path use, and ideal storage allows these copies to be retained until all pairwise preparations are performed.
Therefore
\begin{equation}
W(N)=\max_{e\in E}\ell_e.
\label{eq:gcp-window-identity}
\end{equation}
Equation~\eqref{eq:gcp-window-identity} is an exact finite-graph activation-slot identity for this preparation and assigns no wall-clock duration to an activation slot.

\paragraph{Realizability of the retained effective states.}
For an \(h\)-hop shortest path, call the pure two-qubit state produced between its endpoints by the outcome-corrected deterministic series rule the \emph{path state}; its concurrence is \(\sin^h(2\theta)\)~\cite{meng2021concurrence}.
Define
\begin{equation}
\alpha_h
:=
\frac{1+\sqrt{1-\sin^{2h}(2\theta)}}{2},
\label{eq:gcp-path-eigenvalue}
\end{equation}
which is the larger eigenvalue of either one-qubit reduced density operator of the path state.
For a fixed pair \(\{u,v\}\), all paths in \(\mathcal P_{uv}\) have the same hop length \(h\).
Because each prescribed path is supplied with its own elementary-pair copies, the path states for a fixed pair \(\{u,v\}\) are prepared on separate qubits.
The reduced state at either endpoint is therefore the tensor product of the reduced states of the individual path states, so its largest eigenvalue is \(\alpha_h^{|\mathcal P_{uv}|}\).
The input Schmidt vector is therefore majorized by the two-qubit target Schmidt vector with larger entry
\[
\max\!\left\{\frac12,\alpha_h^{|\mathcal P_{uv}|}\right\}.
\]
Because the target has Schmidt rank two, the bound on the largest Schmidt coefficient is sufficient for this majorization.
The pure-state majorization criterion therefore gives a deterministic LOCC conversion to that effective two-qubit state~\cite{nielsen1999conditions}.
Its singlet-conversion probability is
\begin{equation}
\min\!\left\{
1,\,
2\left(1-\alpha_h^{|\mathcal P_{uv}|}\right)
\right\},
\label{eq:gcp-effective-scp}
\end{equation}
which is the GCP series-parallel prescription for the corresponding node pair~\cite{nath2026general}.
Operations for different unordered node pairs act on separately assigned systems, so the retained effective states can be prepared jointly.

\begin{figure}[htbp]
\centering
\resizebox{0.6\linewidth}{!}{

\begin{tikzpicture}[
  x=1.2cm, y=1.2cm,
  link/.style={draw=black!45, line width=0.65pt},
  vertex/.style={circle, draw=black!85, line width=0.65pt,
                 minimum size=4.5pt, inner sep=0pt},
  zero/.style={vertex, fill=black!85},
  one/.style={vertex, fill=white},
  coordlabel/.style={anchor=north, xshift=16pt, yshift=-11pt,
                     inner sep=0pt, font=\small}
]

\foreach \i in {-2,...,2} {
  \foreach \j in {-2,...,2} {
    \coordinate (A-\i-\j) at
      ({1.5*(\i+\j)}, {sqrt(3)/2*(\i-\j)});
    \coordinate (B-\i-\j) at
      ({1.5*(\i+\j)+1}, {sqrt(3)/2*(\i-\j)});
  }
}

\foreach \i in {-2,...,2} {
  \foreach \j in {-2,...,2} {
    \draw[link] (B-\i-\j) -- (A-\i-\j);
    \ifnum\i<2\relax
      \pgfmathtruncatemacro{\nexti}{\i+1}
      \draw[link] (B-\i-\j) -- (A-\nexti-\j);
    \fi
    \ifnum\j<2\relax
      \pgfmathtruncatemacro{\nextj}{\j+1}
      \draw[link] (B-\i-\j) -- (A-\i-\nextj);
    \fi
  }
}

\foreach \i in {-2,...,2} {
  \foreach \j in {-2,...,2} {
    \node[zero] at (A-\i-\j) {};
    \node[one] at (B-\i-\j) {};
    \node[coordlabel] at (A-\i-\j) {$(\i,\j,0)$};
  }
}

\node[anchor=north, font=\large] at (0.5,-4.3)
  {$\bullet\;\sigma=0\qquad\circ\;\sigma=1$};

\end{tikzpicture}
}
\caption{Local honeycomb coordinates used in the GCP aggregation lower bound.
Filled and open vertices denote the sublattices \(\sigma=0\) and \(\sigma=1\), respectively, with \((i,j,1)\) adjacent to \((i,j,0)\), \((i+1,j,0)\), and \((i,j+1,0)\).}
\label{fig:honeycomb}
\end{figure}
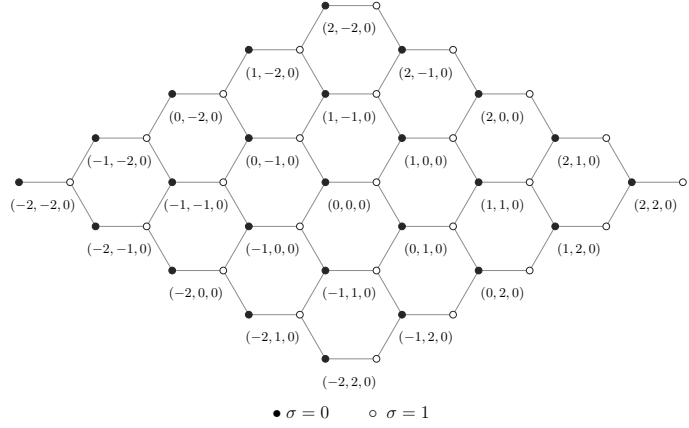

\paragraph{Stretched-exponential aggregation.}
Consider honeycomb coordinates \((i,j,\sigma)\), where \(i,j\in\mathbb Z\), \(\sigma\in\{0,1\}\), and \((i,j,1)\) is adjacent to \((i,j,0)\), \((i+1,j,0)\), and \((i,j+1,0)\), as shown in Figure~\ref{fig:honeycomb}.
The selection of origin \((0,0,0)\) is arbitrary.

Within each resource graph, choose an unperforated coordinate block \(0\le i,j\le m\) with \(m=\Theta(\sqrt N)\).
For the lower bound, consider paths from \((0,0,0)\) to \((m,m,0)\).
The path can be formed by ordering \(m\) increments of the first coordinate and \(m\) increments of the second coordinate, with each increment implemented in two hops through the intermediate sublattice.
This gives
\[
\binom{2m}{m}
\]
distinct paths of length \(4m\), all of which use the link location between \((0,0,0)\) and \((0,0,1)\).
The integer-valued function \((i,j,\sigma)\mapsto2i+2j+\sigma\) changes by one in absolute value across every honeycomb link location and differs by \(4m\) between the two endpoints, so no shorter path exists even if paths outside the coordinate block are allowed.
All of the constructed paths therefore belong to the corresponding shortest-path set, and Eq.~\eqref{eq:gcp-window-identity} gives
\begin{equation}
W(N)
\ge
\binom{2m}{m}
\ge
\frac{4^m}{2m+1}
=
\exp[\Omega(\sqrt N)].
\label{eq:gcp-window-lower}
\end{equation}

For the converse order, every shortest path in the resource graph has \(O(\sqrt N)\) hops because the graph family has hop diameter \(O(\sqrt N)\).
The honeycomb graph has maximum degree three, so the number of nonbacktracking paths of any specified positive length \(h\) from a given node is at most \(3\cdot2^{h-1}\).
Every shortest path is nonbacktracking, and hence
\begin{equation}
W(N)
=
\max_{e\in E}\ell_e
\le
\sum_{\{u,v\}\subseteq V}|\mathcal P_{uv}|
\le
\binom{N}{2}\,3\,2^{O(\sqrt N)}
=
\exp[O(\sqrt N)].
\label{eq:gcp-window-upper}
\end{equation}
Equations~\eqref{eq:gcp-window-lower} and \eqref{eq:gcp-window-upper} prove the second equality in Eq.~\eqref{eq:gcp-window-scaling}.
The lower bound arises from exponentially many prescribed shortest paths concentrating elementary-pair demand on one link location.

\paragraph{Entanglement transport.}
Consider first independent singlet conversion of every retained effective state followed by routing a maximum set of paths that do not reuse any resulting pure Bell pair.
For a neighboring requested pair, the retained state is its dedicated elementary state, whose singlet-conversion probability is \(2\sin^2\theta>0\), and a successful conversion directly delivers one exact Bell pair.
Since the resource graph has \(|E|=\Theta(N)\) nearest-neighbor pairs among \(\binom N2=\Theta(N^2)\) uniformly sampled sessions and fixed positive nearest-neighbor physical spacing,
\begin{equation}
\mathbb E\!\left[
d(s_1,t_1)B_1^{(W(N))}
\right]
\ge
\frac{2\sin^2\theta}{\binom N2}
\sum_{\{u,v\}\in E}d(u,v)
=
\Omega(N^{-1}).
\label{eq:gcp-output-lower}
\end{equation}

For the upper bound, consider arbitrary final LOCC acting only on the retained effective states.
Across the cut separating \(s_1\) from all other nodes, only the \(N-1\) retained two-qubit states involving \(s_1\) are nonlocal, so conditioned on their preparation outcomes the joint Schmidt rank across the cut is at most \(2^{N-1}\).
Decompose each conditional local operation in the final LOCC into single-Kraus branches.
Conditioned on a complete sequence of such branches, the global state remains pure, and local operations cannot increase its Schmidt rank across the endpoint cut.
Because \(F_\star=1\), every pair counted by \(B_1^{(W(N))}\) is an ideal Bell pair and therefore factors from the remaining systems on the branch.
A branch delivering \(B_1^{(W(N))}\) counted pairs consequently has Schmidt rank at least \(2^{B_1^{(W(N))}}\) across the endpoint cut, which gives
\begin{equation}
B_1^{(W(N))}
\le
N-1
\label{eq:gcp-output-count-upper}
\end{equation}
on every branch.
The fixed physical spacing and \(O(\sqrt N)\) hop diameter imply \(d(s_1,t_1)=O(\sqrt N)\), and hence
\begin{equation}
\mathbb E\!\left[
d(s_1,t_1)B_1^{(W(N))}
\right]
=
O(N^{3/2}).
\label{eq:gcp-output-upper}
\end{equation}
Combining Eqs.~\eqref{eq:gcp-output-lower} and \eqref{eq:gcp-output-upper} with \(W(N)=\exp[\Theta(\sqrt N)]\) proves Eq.~\eqref{eq:gcp-final-locc-scaling}, and the lower-bound strategy shows that independent singlet conversion followed by maximum pure-Bell-pair routing attains the same order.
Polynomial factors in the unnormalized delivered transport do not change this stretched-exponential classification.

\paragraph{Preparation of the effective states.}
The window and transport orders in Proposition~\ref{prop:gcp-direct-realization} concern the specified direct preparation and final processing of the retained effective states after all other quantum systems are discarded.
They hold for every fixed \(0<\theta<\pi/4\), including the reported GCP threshold.

On the regular honeycomb graph family, a spatial cut with \(\Theta(N)\) nodes on each side intersects \(O(\sqrt N)\) elementary link locations and separates \(\Theta(N^2)\) effective pairs.
Since each effective pair is pure and entangled, the target product state has Schmidt rank \(2^{\Theta(N^2)}\) across this cut, whereas the resources supplied in \(W\) slots have Schmidt rank at most \(2^{O(W\sqrt N)}\).
Monotonicity of Schmidt rank on every LOCC branch~\cite{terhal2000schmidt} therefore requires \(W=\Omega(N^{3/2})\) for exact simultaneous preparation of the effective states.

A polynomial window suffices under the same resource assumptions.
For fixed \(\theta\), choose an integer \(k\), independent of \(N\), such that \((\cos^2\theta)^k\leq 1/2\); the pure-state majorization criterion then permits deterministic conversion of \(k\) elementary pairs on a link into one Bell pair~\cite{nielsen1999conditions}.
Choose a spanning tree, prepare each required effective pair locally at one endpoint, and teleport one qubit along the tree path to the other endpoint, using a separate Bell pair on each traversed edge.
Each tree edge lies on at most \(\lfloor N^2/4\rfloor\) paths between unordered node pairs, so \(W=k\lfloor N^2/4\rfloor=O(N^2)\) slots suffice.

\section{Derivations for the chain resource-layer benchmark}
\label{app:chain-benchmark}

We use the chain resource-layer assumptions of Sec.~\ref{sec:chain-benchmark}, including random exclusive pairing, ideal resource processing and storage, and access to all heralded realized link information.
The calibration and asymptotic aggregation arguments below allow any fixed \(F_\star\in(F_{\min},1)\); the finite-window comparisons and single-slot identity use \(F_\star=0.95\).

\paragraph{Random-demand identity.}
For any prescribed unordered node pair \(\{s,t\}\), uniform active-user sampling followed by uniform exclusive pairing gives
\[
\Pr[\{s,t\}\in\mathcal D]
=\frac{M(M-1)}{N(N-1)}\frac{1}{M-1}
=\frac{M}{N(N-1)}.
\]
There are \(N-h\) unordered node pairs at hop distance \(h\), so any function \(f\) satisfies
\begin{equation}
\mathbb E_{\mathcal D}\!\left[\sum_{i=1}^{M/2}f(h_i)\right]
=\frac{M}{N(N-1)}\sum_{h=1}^{N-1}(N-h)f(h).
\label{eq:chain-demand-identity}
\end{equation}
In particular, a session chosen uniformly from \(\mathcal D\) has mean hop distance \((N+1)/3\).

\paragraph{Perfect-output allocation over a finite aggregation window.}
Fix a finite \(W\ge1\) and a demand realization \(\mathcal D\).
For the allocation arguments in this appendix, let \(K_x\) count the elementary pairs supplied at link location \(x\), between nodes \(x\) and \(x+1\), over the aggregation window.
Then \(K_x\sim\operatorname{Binomial}(wW,p)\), independently over \(x=0,\ldots,N-2\).
Define the local supply vector and incidence matrix by
\[
\mathbf K=(K_0,\ldots,K_{N-2})^\top,
\qquad
A_{xi}=\mathbf 1\{s_i\le x<t_i\},
\qquad
\mathbf A\in\{0,1\}^{(N-1)\times M/2}.
\]
For the cut converse, every admissible strategy can be simulated using all elementary pairs supplied within the aggregation window as its initial resource.
Across cut \(x\), this resource has Schmidt number \(2^{K_x}\), which cannot increase under any conditional LOCC outcome~\cite{terhal2000schmidt}.
At \(F_\star=1\), jointly counted Bell pairs are pure and therefore factor from one another and from the remaining output systems.
Consequently, their integer counts \(\mathbf b=(b_i)_{i=1}^{M/2}\) satisfy \(\mathbf A\mathbf b\le\mathbf K\), with vector inequalities interpreted componentwise.
Conversely, every nonnegative integer vector satisfying these constraints can be realized by assigning distinct elementary pairs along the session paths and performing ideal swapping.
Ideal storage permits this allocation after all link opportunity outcomes within the aggregation window are known.
Thus Eq.~\eqref{eq:resource-et-capacity}, evaluated at \(F_\star=1\), gives the identity
\begin{equation}
C_{\mathrm{ET}}^{(W)}(N,M;1)
=\frac{1}{W}\mathbb E_{\mathcal D,\mathbf K}\!\left[
\max_{\substack{\mathbf b\in\mathbb Z_{\ge0}^{M/2}\\\mathbf A\mathbf b\le\mathbf K}}
a_N\sum_{i=1}^{M/2}h_i b_i
\right].
\label{eq:app-chain-window-allocation}
\end{equation}
Each column of \(\mathbf A\) has consecutive ones, so \(\mathbf A\) is totally unimodular and, because \(\mathbf K\) is integral, relaxing \(\mathbf b\) to nonnegative real values leaves the optimum unchanged~\cite{fulkerson1965incidence}.

\paragraph{Fidelity-calibrated cut constraints.}
Fix \(F_\star\in(F_{\min},1)\), a finite \(W\ge1\), and an admissible strategy.
For fixed \(\mathcal D\) and heralded realized link information, the supplied elementary pairs contain exactly \(K_x\) units of squashed entanglement across chain cut \(x\).
Applying Eq.~\eqref{eq:app-fidelity-general-cut} and then averaging over realized link histories with the same \(\mathbf K\) gives
\begin{equation}
\sum_{i:s_i\le x<t_i}
\mathbb E[B_i^{(W)}\mid\mathcal D,\mathbf K]
\le\kappa(F_\star)K_x,
\qquad x=0,\ldots,N-2.
\label{eq:app-fidelity-cut-count}
\end{equation}
Dividing these conditional expected counts by \(\kappa(F_\star)\) gives a feasible fractional allocation for Eq.~\eqref{eq:app-chain-window-allocation}.
By the integrality established above, its objective is bounded by the integer optimum.
Averaging over \(\mathcal D,\mathbf K\), dividing by \(W\), and taking the supremum over admissible strategies proves the upper bound in Eq.~\eqref{eq:chain-window-fidelity-comparison} at \(F_\star=0.95\).
The ideal Bell pairs produced by the allocation construction satisfy this threshold and give the lower bound.
The comparison holds for every finite \(W\), including \(W\) depending on \(N\).
Averaging Eq.~\eqref{eq:app-fidelity-cut-count} over \(\mathbf K\) at fixed \(\mathcal D\), using \(\mathbb E[K_x\mid\mathcal D]=Wwp\), also gives
\begin{equation}
\sum_{i:s_i\le x<t_i}
\frac{\mathbb E[B_i^{(W)}\mid\mathcal D]}{W}
\le\kappa(F_\star)wp,
\qquad x=0,\ldots,N-2.
\label{eq:app-fidelity-calibrated-cuts}
\end{equation}

\paragraph{Fixed-window asymptotics and pre-heralding achievability.}
First set \(W=1\), with \(w\) and \(p\in(0,1)\) fixed as \(N\to\infty\).
All asymptotic estimates in this single-slot calculation are uniform over even \(2\le M\le N\).
For the single-slot derivations, let \(\mu_h\) denote the expected minimum of \(h\) independent copies of \(K_x\).
With the binomial tails \(q_m\), the tail-sum identity gives
\[
\begin{aligned}
q_m&:=\Pr[K_x\ge m]=\sum_{k=m}^{w}\binom{w}{k}p^k(1-p)^{w-k},
\qquad m=1,\ldots,w,\\
\mu_h&=\sum_{m=1}^{w}q_m^h.
\end{aligned}
\]
Ignoring competition among sessions bounds Eq.~\eqref{eq:app-chain-window-allocation} by
\[
C_{\mathrm{ET}}^{(1)}(N,M;1)
\le\overline C_{\mathrm{bott}}^{(1)}
:=\frac{Ma_N}{N(N-1)}\sum_{h=1}^{N-1}(N-h)h\mu_h,
\]
where \(\overline C_{\mathrm{bott}}^{(1)}\) is an auxiliary bound used only in this proof.
Since \(0<q_m<1\), the geometric series and dominated convergence give
\[
\Gamma_{w,p}:=\sum_{h\ge1}h\mu_h
=\sum_{m=1}^{w}\frac{q_m}{(1-q_m)^2}\in(0,\infty),
\qquad
\frac{N\overline C_{\mathrm{bott}}^{(1)}}{Ma_N}\longrightarrow\Gamma_{w,p},\qquad N\to\infty.
\]

For a constructive lower bound, consider a pre-heralding strategy that, for each demand realization, selects pairwise link-disjoint sessions before observing the link opportunity outcomes and assigns all \(w\) link opportunities at each link location along every selected path.
Ideal swapping delivers the minimum number of supplied elementary pairs along each selected path, and all delivered outputs are perfect Bell pairs.
Its expected entanglement transport per activation slot is therefore
\[
\underline C_{\mathrm{pre}}^{(1)}
=\mathbb E_{\mathcal D}\!\left[
\max_{\substack{\mathbf z\in\{0,1\}^{M/2}\\\mathbf A\mathbf z\le\mathbf 1}}
a_N\sum_{i=1}^{M/2}h_i\mu_{h_i}z_i
\right],
\]
where \(z_i\in\{0,1\}\) indicates whether session \(i\) is selected.
For the following estimate, set \(v_h=h\mu_h\), \(q=q_1<1\), \(B=\sup_{h\ge1}v_h<\infty\), and \(H_N=\lceil\log N\rceil\).
Because \(v_h\le whq^h\), Eq.~\eqref{eq:chain-demand-identity} gives
\[
\mathbb E_{\mathcal D}\!\left[\sum_{i:h_i>H_N}a_Nv_{h_i}\right]
\le\frac{Ma_N}{N-1}\sum_{h>H_N}v_h
=o\!\left(\frac{Ma_N}{N}\right).
\]
Among sessions with \(h_i\le H_N\), retain those sharing no link location with any other such session.
There are at most \(2NH_N^3\) candidate pairs of overlapping intervals of length at most \(H_N\): at most \(N\) choices for the first starting node, \(H_N^2\) choices for the lengths, and \(2H_N\) choices for the second starting node.
For \(N\ge4\), two prescribed intervals with four distinct endpoints both belong to \(\mathcal D\) with probability
\[
\frac{M(M-2)}{N(N-1)(N-2)(N-3)},
\]
whereas intervals sharing an endpoint cannot both occur.
If the local variable \(Z_N\) counts overlapping candidate pairs present in \(\mathcal D\), then
\[
\mathbb E[Z_N]=O\!\left(\frac{M^2H_N^3}{N^3}\right),
\qquad
2a_NB\,\mathbb E[Z_N]
=o\!\left(\frac{Ma_N}{N}\right),
\]
because \(M\le N\) and \(H_N^3/N\to0\).
At most \(2Z_N\) short sessions are discarded, so their expected contribution to the auxiliary bound is at most \(2a_NB\,\mathbb E[Z_N]\).
The maximizing pre-heralding selection performs at least as well as this feasible selection, so the long-session and overlap estimates give \(\overline C_{\mathrm{bott}}^{(1)}-\underline C_{\mathrm{pre}}^{(1)}=o(Ma_N/N)\).
Since \(\underline C_{\mathrm{pre}}^{(1)}\le C_{\mathrm{ET}}^{(1)}(N,M;1)\le\overline C_{\mathrm{bott}}^{(1)}\), the preceding bounds give
\begin{equation}
\frac{N}{Ma_N}\underline C_{\mathrm{pre}}^{(1)}
\longrightarrow \Gamma_{w,p},
\qquad
\frac{N}{Ma_N}C_{\mathrm{ET}}^{(1)}(N,M;1)
\longrightarrow \Gamma_{w,p},
\qquad N\to\infty,
\label{eq:app-single-slot-asymptotic}
\end{equation}
uniformly over even \(2\le M\le N\). For \(w=1\), \(\mu_h=p^h\), so \(\Gamma_{1,p}=p/(1-p)^2\).

For general finite \(W\), define \(\Gamma_{wW,p}\) by the preceding formula for \(\Gamma_{w,p}\) with \(w\) replaced by \(wW\).
The same bottleneck and pre-heralding arguments apply to the accumulated \(\operatorname{Binomial}(wW,p)\) supply, with the selected sessions fixed before any link opportunity outcome in the aggregation window is observed.
For fixed \(w\), \(p\in(0,1)\), and \(W\) independent of \(N\), division by \(W\) therefore proves Eq.~\eqref{eq:chain-window-perfect-reference} as \(N\to\infty\), uniformly over even \(2\le M\le N\).
Combining this limit with Eq.~\eqref{eq:chain-window-fidelity-comparison} proves Theorem~\ref{thm:chain-single-slot} at \(F_\star=0.95\).
The coefficient \(\Gamma_{wW,p}/W\) is finite and independent of \(N,M\) under these fixed-parameter assumptions and belongs to the entanglement transport capacity at \(F_\star=1\); the calibrated bounds do not determine the leading coefficient at \(F_\star=0.95\) or establish attainability of \(\kappa(0.95)\).
This normalized limit does not cover \(W\) growing with \(N\).

\paragraph{Exact single-slot identity.}
Set \(w=1\), \(W=1\), and \(F_\star=0.95\), retaining access to all heralded realized link information.
For fixed \(\mathcal D\) and slot realization, each cut has \(K_x\in\{0,1\}\).
If \(K_x=0\), the input is separable across the cut, so every crossing two-qubit output state has target-Bell fidelity at most \(1/2\).
If \(K_x=1\), every conditional LOCC output has Schmidt number at most two across the cut, including after local systems are discarded~\cite{terhal2000schmidt}.
Suppose two fidelity-qualified Bell pairs cross this cut, and denote their joint conditional state by \(\rho\) and the target projector by \(P=\proj{\Phi^+}\), locally to this proof.
Positivity of \((\I-P)\otimes(\I-P)\) gives
\[
\tr[(P\otimes P)\rho]
\ge\tr[(P\otimes\I)\rho]+\tr[(\I\otimes P)\rho]-1
\ge2F_\star-1=0.90>\frac12.
\]
However, \(P\otimes P\) projects onto a maximally entangled state of local dimension four across the cut, whose overlap with a state of Schmidt number at most two is at most \(2/4=1/2\)~\cite{terhal2000schmidt}.
Thus the counts \(b_i\) of fidelity-qualified Bell pairs satisfy \(\mathbf A\mathbf b\le\mathbf K\) for every conditional output.
Equation~\eqref{eq:app-chain-window-allocation} and ideal swapping then give \(C_{\mathrm{ET}}^{(1)}(N,M;0.95)=C_{\mathrm{ET}}^{(1)}(N,M;1)\), proving Proposition~\ref{prop:fq-single-slot}.

\paragraph{Asymptotic aggregation.}
Fix \(N\), \(M\), \(w\), \(p\in(0,1)\), and \(F_\star\in(F_{\min},1)\).
Choose one session uniformly from \(\mathcal D\), retain its elementary pairs throughout the aggregation window, and perform ideal swapping to deliver the minimum accumulated count along its path.
For each fixed demand realization and chosen session \(i\), the law of large numbers and \(0\le K_x/W\le w\) imply
\[
\lim_{W\to\infty}\frac{1}{W}
\mathbb E\!\left[\min_{s_i\le x<t_i}K_x\,\middle|\,\mathcal D,i\right]
=wp.
\]
Averaging the session distance gives the lower bound below, while summing Eq.~\eqref{eq:app-fidelity-calibrated-cuts} over the \(N-1\) cuts gives the upper bound, since \(\sum_x A_{xi}=h_i\).
Using Eq.~\eqref{eq:asymptotic-capacity},
\[
wp\,a_N\frac{N+1}{3}
\le C_{\mathrm{ET}}^{(\infty)}(N,M;F_\star)
\le\kappa(F_\star)wp\,a_N(N-1).
\]
Hence \(C_{\mathrm{ET}}^{(\infty)}(N,M;F_\star)=\Theta(wp\,a_NN)\) as \(N\to\infty\), for fixed \(w\), \(p\in(0,1)\), and \(F_\star\in(F_{\min},1)\), uniformly over even \(2\le M\le N\).
At \(F_\star=0.95\), this proves Theorem~\ref{thm:chain-asymptotic} and, together with Theorem~\ref{thm:chain-single-slot}, Eq.~\eqref{eq:chain-aggregation-separation} for fixed \(W\).

The corresponding ideal-ebit cut/flow reference for asymptotic aggregation is
\begin{equation}
\begin{aligned}
\overline C_{\mathrm{ET}}^{(\infty)}(\mathcal D;1)
:=\max_{\mathbf R^{\mathrm{ebit}}\ge0}\quad
&a_N\sum_{i=1}^{M/2}h_iR_i^{\mathrm{ebit}}\\
\text{subject to}\quad
&\sum_{i:s_i\le x<t_i}R_i^{\mathrm{ebit}}\le wp,
\qquad x=0,\ldots,N-2,
\end{aligned}
\label{eq:app-chain-ideal-ebit-lp}
\end{equation}
where \(R_i^{\mathrm{ebit}}\) is an ideal Bell-pair rate per activation slot, local to this reference.
After division of the rates by \(wp\), interval-matrix integrality gives a binary optimum selecting pairwise link-disjoint sessions.
Ideal storage and swapping achieve rate \(wp\) on each selected path by the same fixed-\(N\) limit used above, so this ideal-ebit reference is achievable on the chain.
For any finite \(W\), Eq.~\eqref{eq:app-fidelity-calibrated-cuts} also gives
\begin{equation}
\frac{1}{W}\mathbb E\!\left[
\sum_{i=1}^{M/2}a_Nh_iB_i^{(W)}\,\middle|\,\mathcal D
\right]
\le\kappa(F_\star)
\overline C_{\mathrm{ET}}^{(\infty)}(\mathcal D;1).
\label{eq:app-fidelity-window-converse}
\end{equation}

\section{Physical implementations for the chain benchmark}
\label{app:physical-models}

\subsection{Proof of the physical hop-profile scaling law}
\label{app:physical-multiuser}

We use the construction defined in Section~\ref{sec:chain-physical} and the hop profile in Eq.~\eqref{eq:physical-hop-profile}.
Removing the pairwise link-disjoint constraint gives
\[
\underline S_{\mathrm{ET}}(N,M;F_\star)
\le
\mathbb E_{\mathcal D}\!\left[
\sum_{i=1}^{M/2}a_Nh_i b_{h_i}
\right]
=
\frac{Ma_N}{N(N-1)}
\sum_{h=1}^{N-1}(N-h)h b_h,
\]
where Eq.~\eqref{eq:chain-demand-identity} was used.
If \(0<\Lambda<\infty\), dominated convergence gives
\[
\frac{N}{Ma_N}
\mathbb E_{\mathcal D}\!\left[
\sum_{i=1}^{M/2}a_Nh_i b_{h_i}
\right]
=
\frac{1}{N-1}
\sum_{h=1}^{N-1}(N-h)h b_h
\longrightarrow
\Lambda,
\qquad N\to\infty.
\]

For the matching lower bound, set \(H_N=\lceil\log N\rceil\).
The expected entanglement transport contributed by demand sessions with \(h_i>H_N\) satisfies
\[
\begin{aligned}
\mathbb E_{\mathcal D}\!\left[
\sum_{i:h_i>H_N}a_Nh_i b_{h_i}
\right]
&=
\frac{Ma_N}{N(N-1)}
\sum_{h=H_N+1}^{N-1}(N-h)h b_h\\
&\le
\frac{Ma_N}{N-1}
\sum_{h>H_N}h b_h\\
&=
o\!\left(\frac{Ma_N}{N}\right),
\end{aligned}
\]
because \(H_N\to\infty\) and \(\sum_{h\ge1}h b_h=\Lambda<\infty\).
Among the demand sessions with \(h_i\le H_N\), let \(Z_N\) count overlaps between two sessions whose chain paths share at least one link location.
The overlap counting used to prove Eq.~\eqref{eq:app-single-slot-asymptotic} gives
\[
\mathbb E[Z_N]
=
O\!\left(\frac{M^2H_N^3}{N^3}\right).
\]
Deleting every short session involved in an overlap leaves a pairwise link-disjoint set.
Since \(h b_h\le\Lambda\) for every \(h\), the expected service removed by this deletion is at most
\[
2a_N\Lambda\,\mathbb E[Z_N]
=
O\!\left(
\frac{a_N\Lambda M^2H_N^3}{N^3}
\right)
=
o\!\left(\frac{Ma_N}{N}\right)
\]
uniformly over even \(2\le M\le N\), because \(M\le N\) and \(H_N^3/N\to0\).
The maximum-weight selection in the construction performs at least as well as this feasible selection.
Combining the upper and lower bounds proves
\begin{equation}
\frac{N}{Ma_N}\underline S_{\mathrm{ET}}(N,M;F_\star)
\longrightarrow
\Lambda,
\qquad N\to\infty,
\qquad
0<\Lambda<\infty,
\label{eq:app-physical-chain-summable}
\end{equation}
uniformly over even \(2\le M\le N\), with the expansion spacing and all physical parameters fixed.
Thus, under the construction and concurrency assumptions of Theorem~\ref{thm:hop-profile}, a finite hop ceiling is sufficient but not necessary; the proof uses only \(0<\Lambda<\infty\).

\paragraph{Single-slot reference ratio.}
At the matched benchmark \(w=W=1\), \(F_\star=0.95\), with access to all heralded realized link information, Proposition~\ref{prop:fq-single-slot} gives
\[
C_{\mathrm{ET}}^{(1)}(N,M;0.95)
=
C_{\mathrm{ET}}^{(1)}(N,M;1).
\]
At \(w=1\), Eq.~\eqref{eq:app-single-slot-asymptotic} gives
\[
\frac{N}{Ma_N}C_{\mathrm{ET}}^{(1)}(N,M;1)
\longrightarrow
\Gamma_{1,p},
\qquad N\to\infty,
\qquad
\Gamma_{1,p}=\frac{p}{(1-p)^2}.
\]
Combining the exact identity above with Eq.~\eqref{eq:app-physical-chain-summable} gives
\begin{equation}
\frac{\underline S_{\mathrm{ET}}(N,M;0.95)}
{rC_{\mathrm{ET}}^{(1)}(N,M;0.95)}
\longrightarrow
\frac{\Lambda}{r\Gamma_{1,p}}
=
\frac{\Lambda(1-p)^2}{rp},
\qquad N\to\infty.
\label{eq:app-physical-single-slot-ratio}
\end{equation}
The identity of the two single-slot capacities is exact at this benchmark point, whereas Eq.~\eqref{eq:app-physical-single-slot-ratio} is a large-\(N\) limit.

\subsection{Matched noisy physical implementations}
\label{app:matched-noisy-hardware}

\paragraph{Physical parameters and timing assumptions.}
The matched benchmark uses the initialization, memory-dephasing, and effective swapping-error maps specified below, drawing on Kinos et al.~\cite{kinos2025complete}, while the link opportunity probability, retention, timing, and provisioned parallel operation are specified here.
We use \(a_N=a=1\,\mathrm{km}\), \(w=1\), and \(F_\star=0.95\).
The two-photon success law follows Eqs.~(3)--(4) of Ref.~\cite{kinos2025complete}, with its distance-independent prefactor fixed here at \(0.05\):
\[
p=0.05\exp\!\left[-\frac{a_N}{22\,\mathrm{km}}\right]
\simeq0.0477781518.
\]
At each active link location, the corresponding link opportunity is attempted independently once per reference interval
\[
1/r=45\,\mu\mathrm{s},
\qquad
r\simeq2.22222222\times10^4\,\mathrm{s}^{-1}.
\]
The interval consists of an allocated \(40\,\mu\mathrm{s}\) local preparation and generation budget followed by \(5\,\mu\mathrm{s}\) for propagation and heralding.
At a fiber propagation speed of \(2\times10^8\,\mathrm{m}/\mathrm{s}\), propagation from a node to the midpoint detector and return of the herald takes \(5\,\mu\mathrm{s}\) for \(a=1\,\mathrm{km}\).
Kinos et al. obtain a \(40\,\mu\mathrm{s}\) interval between attempts at the same link location by doubling approximately \(10\,\mu\mathrm{s}\) of initialization plus \(10\,\mu\mathrm{s}\) of generation because a shared communication ion alternates between neighboring link locations~\cite[Sec.~IV, p.~033199-8]{kinos2025complete}.
The present benchmark instead provisions independent communication resources for parallel operation at neighboring link locations and retains \(40\,\mu\mathrm{s}\) as a conservative local budget.
Local work and the final propagation interval are serial, with no overlapping attempts at the same link location.
Neighboring-node purification-outcome exchange separately takes \(5\,\mu\mathrm{s}\), following the separation of local operations and propagation in Eq.~(6) of Ref.~\cite{kinos2025complete}.
We take
\[
T_{\mathrm{coh}}=0.5\,\mathrm{s},
\qquad
\epsilon_{\mathrm{i}}
=\epsilon_{\mathrm{g}}
=\epsilon_{\mathrm{m}}
=10^{-3},
\qquad
t_{\mathrm{swap}}=210\,\mu\mathrm{s},
\qquad
t_{\mathrm{pur}}=220\,\mu\mathrm{s}.
\]
The operation-time budgets are taken from Sec.~IV of Ref.~\cite{kinos2025complete}.
Here \(\epsilon_{\mathrm{i}}\) and \(\epsilon_{\mathrm{m}}\) are initialization and measurement error probabilities.
The parameter \(\epsilon_{\mathrm{g}}\) specifies an effective Bell-state error for swapping and a total local two-qubit Pauli-error probability for each DEJMPS controlled-NOT, according to the separate maps below; using the same numerical value does not identify a common microscopic gate channel.
The retention limits count stored elementary pairs per link location and exclude the temporary memory required for a fresh attempt.
Fresh attempts are assumed to leave retained elementary pairs unaffected except for their specified memory dephasing.
Selected sessions operate concurrently with independent link opportunity outcomes and sufficient local resources, so their isolated-session rates are unchanged by concurrent operation.
Path-wide signaling delays are neglected as specified in Sec.~\ref{sec:chain-physical}.
For \(h\ge2\), a fidelity-qualified Bell pair is counted at swap completion, conditioned on the swapping outcomes; for \(h=1\), it is counted when the successful-attempt herald arrives.
Endpoint buffering after counting is outside the physical implementation.

\paragraph{Bell-diagonal state evolution.}
We represent every retained elementary pair by Bell-state probabilities
\[
\mathbf c=(A,B,C,D)
\]
in the ordered basis
\[
\left(
\ket{\Phi^+},
\ket{\Psi^-},
\ket{\Psi^+},
\ket{\Phi^-}
\right).
\]
The framework uses \(\ket{\Phi^+}\) as its canonical target Bell state.
Since \(\ket{\Psi^+}=(\I\otimes X)\ket{\Phi^+}\), the two conventions differ only by a fixed local Pauli transformation.
Thus \(C\ge F_\star\) is equivalent to the framework fidelity qualification after that fixed local correction.

For an elementary pair, let \(t=0\) denote herald arrival at the end of its successful activation slot.
The effective initialization and storage model assigns initialization errors before that activation slot and independent memory dephasing over exactly one complete reference interval \(1/r\).
Accordingly, the pair is dephased over one full reference interval before \(t=0\); this is imposed at the Bell-state level rather than derived from the microscopic preparation sequence.
The resulting coefficients at herald arrival are
\begin{align}
A(0)&=D(0)=0,\nn\\
B(0)&=\frac{1}{2}\left[1-(1-2\epsilon_{\mathrm{i}})^2e^{-2/(rT_{\mathrm{coh}})}\right],\nn\\
C(0)&=\frac{1}{2}\left[1+(1-2\epsilon_{\mathrm{i}})^2e^{-2/(rT_{\mathrm{coh}})}\right].
\label{eq:physical-initial-bell}
\end{align}
For Bell-diagonal coefficients \((A,B,C,D)\) at the beginning of any subsequent storage interval of physical duration \(t\), independent phase decoherence of the two constituent memories gives
\begin{align}
A(t)&=\frac{A+D}{2}+\frac{A-D}{2}e^{-2t/T_{\mathrm{coh}}},\nn\\
B(t)&=\frac{B+C}{2}+\frac{B-C}{2}e^{-2t/T_{\mathrm{coh}}},\nn\\
C(t)&=\frac{C+B}{2}+\frac{C-B}{2}e^{-2t/T_{\mathrm{coh}}},\nn\\
D(t)&=\frac{D+A}{2}+\frac{D-A}{2}e^{-2t/T_{\mathrm{coh}}}.
\label{eq:physical-dephasing-map}
\end{align}
The factor \(e^{-2t/T_{\mathrm{coh}}}\) results from independent dephasing of the two constituent memories during the same physical interval \(t\); it does not represent a doubled wall-clock duration.

For Bell-diagonal inputs \(\mathbf c_L=(A_L,B_L,C_L,D_L)\) and \(\mathbf c_R=(A_R,B_R,C_R,D_R)\), ideal deterministic swapping gives~\cite{kinos2025complete}
\begin{align}
A'&=A_LC_R+C_LA_R+B_LD_R+D_LB_R,\nn\\
B'&=A_LD_R+D_LA_R+B_LC_R+C_LB_R,\nn\\
C'&=A_LA_R+B_LB_R+C_LC_R+D_LD_R,\nn\\
D'&=A_LB_R+B_LA_R+C_LD_R+D_LC_R.
\label{eq:physical-swap-map}
\end{align}
Each swap has an effective Bell-state error probability \(\epsilon_{\mathrm{g}}\), assigning probability \(\epsilon_{\mathrm{g}}/3\) to each of the other three Bell states.
After the \(h-1\) swaps of an \(h\)-hop path, this map acts on each Bell coefficient \(X\in\{A,B,C,D\}\) as
\begin{equation}
X
\longmapsto
\frac14+
\left(X-\frac14\right)
\left(1-\frac{4\epsilon_{\mathrm{g}}}{3}\right)^{h-1}.
\label{eq:physical-gate-map}
\end{equation}
For the measurement errors, define the local quantity
\[
e_h
:=
\frac{1+(1-2\epsilon_{\mathrm{m}})^{h-1}}{2}.
\]
The resulting erroneous Pauli corrections act on the Bell coefficients as
\begin{align}
A&\longmapsto e_h^2A+(1-e_h)e_hD+e_h(1-e_h)C+(1-e_h)^2B,\nn\\
B&\longmapsto e_h^2B+(1-e_h)e_hC+e_h(1-e_h)D+(1-e_h)^2A,\nn\\
C&\longmapsto e_h^2C+(1-e_h)e_hB+e_h(1-e_h)A+(1-e_h)^2D,\nn\\
D&\longmapsto e_h^2D+(1-e_h)e_hA+e_h(1-e_h)B+(1-e_h)^2C.
\label{eq:physical-measurement-map}
\end{align}
Equations~\eqref{eq:physical-initial-bell}--\eqref{eq:physical-measurement-map} define the common noisy state evolution used by the two physical implementations below, with subsequent storage times measured from herald arrival.

\paragraph{Noisy physical implementation.}
For an isolated \(h\)-hop session, activation slots of a delivery cycle are indexed by \(\tau=1,2,\ldots\) from the start of the cycle, and the link opportunity at every link location is attempted independently once per reference interval \(1/r\).
A successful attempt in activation slot \(\tau\) is heralded at time \(\tau/r\).
Let \(\tau_{\mathrm{ready}}\) be the first activation slot at whose end every link location has supplied at least one elementary pair.
Each link location retains its newest elementary pair while waiting for the remaining link locations; let \(\tau_j\le\tau_{\mathrm{ready}}\) be the heralding activation slot of the retained elementary pair at link location \(j\).
For \(h=1\), the first successful herald completes the delivery cycle at time \(\tau_{\mathrm{ready}}/r\), without swapping or purification.
For \(h\ge2\), generation stops at time \(\tau_{\mathrm{ready}}/r\), and all \(h-1\) swapping operations run in parallel for \(t_{\mathrm{swap}}\).
The successful activation slot has already been included in Eq.~\eqref{eq:physical-initial-bell}, so the retained elementary pair at link location \(j\) undergoes subsequent storage for
\[
(\tau_{\mathrm{ready}}-\tau_j)/r+t_{\mathrm{swap}}.
\]
The dephasing over \(t_{\mathrm{swap}}\) is included in this storage before the elementary-pair states are combined by Eq.~\eqref{eq:physical-swap-map}, after which the swapping-error maps in Eqs.~\eqref{eq:physical-gate-map} and~\eqref{eq:physical-measurement-map} are applied.
This ordering specifies the placement of storage noise relative to the instantaneous state maps.
The output is counted if and only if its conditional fidelity satisfies \(C\ge F_\star\).
For \(h\ge2\), the delivery-cycle duration is
\[
T_{\mathrm{cyc}}=\tau_{\mathrm{ready}}/r+t_{\mathrm{swap}}.
\]
Every cycle discards any remaining elementary pairs and restarts generation, regardless of output qualification; preparation for a fresh attempt is included in the next reference interval.

\paragraph{One-round DEJMPS physical implementation.}
This physical implementation uses the same link opportunity probability, dephasing, and swapping maps but allows each link location to retain its three most recent elementary pairs.
At time \(\tau_{\mathrm{ready}}/r\), a link location with three retained elementary pairs designates its two newest elementary pairs as DEJMPS inputs and its third-newest elementary pair as fallback, while a link location with fewer than three retained elementary pairs designates its newest elementary pair for direct use.
If no link location is eligible for purification, the cycle follows the noisy physical implementation without purification overhead.
Otherwise, generation remains stopped while all eligible link locations attempt one DEJMPS round in parallel over \(t_{\mathrm{pur}}\).
A DEJMPS input heralded in activation slot \(\tau\) is dephased for
\[
(\tau_{\mathrm{ready}}-\tau)/r+t_{\mathrm{pur}}
\]
before the instantaneous noisy circuit is evaluated.
The circuit applies the local \(\pm\pi/2\) rotations and bilateral controlled-NOT gates of DEJMPS, followed by computational-basis measurements of the sacrificial elementary pair~\cite{kinos2025complete}.
After each local controlled-NOT, each of the 15 nonidentity two-qubit Pauli operators is applied with probability \(\epsilon_{\mathrm{g}}/15\), and the identity is applied with probability \(1-\epsilon_{\mathrm{g}}\).
Each reported measurement bit is independently flipped with probability \(\epsilon_{\mathrm{m}}\).
The purification output is accepted when the two reported outcomes agree, and its state is conditioned on this acceptance event.
The neighboring nodes then exchange their outcomes over \(5\,\mu\mathrm{s}\), after which all swapping operations run in parallel for \(t_{\mathrm{swap}}\).
An accepted purified elementary pair is dephased after the circuit for
\[
5\,\mu\mathrm{s}+t_{\mathrm{swap}}.
\]
If purification fails, the fallback elementary pair heralded in activation slot \(\tau\) remains stored for
\[
(\tau_{\mathrm{ready}}-\tau)/r+t_{\mathrm{pur}}+5\,\mu\mathrm{s}+t_{\mathrm{swap}}.
\]
A directly used elementary pair at a link location that does not purify follows the same waiting time whenever another link location purifies.
When at least one link location attempts purification, the delivery-cycle duration is
\[
T_{\mathrm{cyc}}=\tau_{\mathrm{ready}}/r+t_{\mathrm{pur}}+5\,\mu\mathrm{s}+t_{\mathrm{swap}}.
\]
The selected elementary-pair states then undergo the common swapping maps and individual-output fidelity qualification, and all remaining elementary pairs are discarded before the next cycle.
This retention rule, circuit-level error map, and timing are specified for the present benchmark; they are not identified with the complete fixed-session calculation of Kinos et al.~\cite{kinos2025complete}.

\paragraph{Deterministic fidelity ceilings.}
For the noisy physical implementation, the maximum fidelity at fixed \(h\ge2\) occurs when every retained elementary pair is heralded in activation slot \(\tau_{\mathrm{ready}}\), after which it dephases over \(t_{\mathrm{swap}}\) before the swapping-error maps are applied.
The resulting maximum fidelities satisfy
\[
F_{\max}^{\mathrm{no\,pur}}(9)\simeq0.95502854,
\qquad
F_{\max}^{\mathrm{no\,pur}}(10)\simeq0.94992059<F_\star.
\]
For the one-round DEJMPS physical implementation, at least one link location first supplies an elementary pair in activation slot \(\tau_{\mathrm{ready}}\), so that link location has only one retained elementary pair at time \(\tau_{\mathrm{ready}}/r\) and must use it directly.
Consequently, not every link location can purify.
Before purification, each DEJMPS input has \(A=D=0\), and its coherence \(C-B\) is bounded by its value after the minimum allowed storage; the second-newest input must additionally be at least one reference interval older than the newest input.
Relaxing the allowed input coherences to the rectangle determined by these bounds gives an upper bound over all retained histories.
Each of the accepted output's three quantities \(C+B-A-D\), \(C+A-B-D\), and \(C+D-A-B\) is a ratio of bilinear functions of the input coherences with positive denominator, so its absolute maximum on that rectangle occurs at a corner.
These quantities multiply under ideal swapping, and the final error maps attach nonnegative factors.
Bounding purified branches this way, bounding fallback and direct branches by the corresponding directly used elementary pairs, and retaining at least one direct branch therefore gives an upper bound on the final fidelity for each possible number of purified link locations.
The resulting upper envelope is nonincreasing with \(h\), because removing a link location while retaining a direct branch removes factors of absolute magnitude at most one and reduces the accumulated swapping noise.
At the relevant boundary hop counts, the bounds are attained by a positive-probability history with \(\tau_{\mathrm{ready}}=3\): \(h-1\) link locations supply elementary pairs in activation slots \(1,2,3\), the remaining link location first supplies an elementary pair in activation slot \(3\), and every purification attempt is accepted.
This gives
\[
F_{\max}^{\mathrm{DEJMPS}}(12)\simeq0.95191844,
\qquad
F_{\max}^{\mathrm{DEJMPS}}(13)\simeq0.94795960<F_\star.
\]
Consequently,
\[
b_h^{\mathrm{no\,pur}}=0\quad\text{for all }h\ge10,
\qquad
b_h^{\mathrm{DEJMPS}}=0\quad\text{for all }h\ge13.
\]
These zeros follow from deterministic fidelity bounds, independently of Monte Carlo observations.

\paragraph{Waiting budget.}
For the noisy physical implementation, every elementary pair has \(A=D=0\) at herald arrival, and Eq.~\eqref{eq:physical-dephasing-map} preserves this property while multiplying \(C-B\) by \(e^{-2t/T_{\mathrm{coh}}}\).
By Eq.~\eqref{eq:physical-swap-map}, ideal swapping of such pairs preserves \(A=D=0\) and multiplies their values of \(C-B\), so before the swapping-error maps the delivered pair has
\[
C-B=[C(0)-B(0)]^h\exp\!\left\{-\frac{2}{T_{\mathrm{coh}}}\sum_{j=1}^{h}\left[\frac{\tau_{\mathrm{ready}}-\tau_j}{r}+t_{\mathrm{swap}}\right]\right\}.
\]
Equations~\eqref{eq:physical-gate-map} and~\eqref{eq:physical-measurement-map} are affine, so the delivered fidelity is \(\alpha_h\) plus \(\beta_h\) times this value of \(C-B\), where
\[
\alpha_h=\frac14\left[1+\left(1-\frac{4\epsilon_{\mathrm{g}}}{3}\right)^{h-1}(2e_h-1)\right],
\qquad
\beta_h=\frac12\left(1-\frac{4\epsilon_{\mathrm{g}}}{3}\right)^{h-1}e_h(2e_h-1).
\]
Since \(\alpha_h\le1/2<F_\star\) and \(\beta_h>0\), the delivered pair is fidelity-qualified if and only if
\[
\sum_{j=1}^{h}(\tau_{\mathrm{ready}}-\tau_j)\le\nu_h,
\qquad
\nu_h:=r\left[\frac{T_{\mathrm{coh}}}{2}\ln\frac{\beta_h[C(0)-B(0)]^h}{F_\star-\alpha_h}-h\,t_{\mathrm{swap}}\right],
\]
where the waiting budget \(\nu_h\) is measured in activation slots.
Substituting Eq.~\eqref{eq:physical-initial-bell} gives
\[
\nu_h=\frac{rT_{\mathrm{coh}}}{2}\ln\frac{\beta_h(1-2\epsilon_{\mathrm{i}})^{2h}}{F_\star-\alpha_h}-h\,(1+rt_{\mathrm{swap}}),
\]
so \(\nu_h\) is affine in \(T_{\mathrm{coh}}\), with positive slope if and only if \(\alpha_h+\beta_h(1-2\epsilon_{\mathrm{i}})^{2h}>F_\star\), which holds for \(h\le10\) at the benchmark parameters.
At the benchmark parameters, the reproducibility package gives
\[
\begin{array}{c|ccccccccc}
h & 2 & 3 & 4 & 5 & 6 & 7 & 8 & 9 & 10\\\hline
\nu_h & 495.2 & 432.9 & 370.8 & 308.7 & 246.6 & 184.7 & 122.7 & 60.9 & -1.0
\end{array}
\]
in \texttt{results/waiting\_budget.csv}, and \(\nu_{10}<0\) reproduces the hop ceiling \(9\).
This reduction relies on Bell-diagonal state maps in which storage enters only through memory dephasing; the one-round DEJMPS physical implementation has no single-parameter waiting budget because purification depends on the storage time of each input separately.

\subsection{Numerical evaluation}
\label{app:physical-numerics}

\paragraph{Estimator and simulation provenance.}
For each \(h=2,\ldots,12\), we sample independent delivery cycles directly from geometric link-opportunity success histories, recording each success by its heralding activation slot.
The same success histories are used for both physical implementations, and purification acceptance is sampled from the circuit's conditional success probability.
The retained heralding activation slots determine \(\tau_{\mathrm{ready}}\), each selected branch's storage time, and the delivery-cycle duration under the chronology above.
An output contributes one to the reward only when its individual conditional fidelity satisfies \(C\ge F_\star\).
Each hop uses four independent replicas of \(250000\) shared-history cycles, giving \(10^6\) cycles per hop and \(11\) million cycles in total for each physical implementation.
The reference calculation uses NumPy PCG64 with master seed \(20260905\); seeds are assigned by hop and replica, with separate streams for first successes, subsequent retained-success histories, and purification draws, as recorded in the reproducibility package.
Different hops use independent streams.
Replica results are pooled by total reward and total cycle duration.

For cycle \(\ell=1,\ldots,n_h\), define the local simulation variables
\[
X_{h,\ell}
:=
\mathbf 1\{F_{h,\ell}\ge F_\star\},
\qquad
T_{h,\ell}
:=
\text{physical duration of the delivery cycle}.
\]
The regenerative renewal-reward estimator of \(b_h\) is
\begin{equation}
\widehat b_h
=
\frac{\sum_{\ell=1}^{n_h}X_{h,\ell}}
{\sum_{\ell=1}^{n_h}T_{h,\ell}}
=
\frac{\overline X_h}{\overline T_h}.
\label{eq:physical-rate-estimator}
\end{equation}
For the reported positive-rate rows, define
\[
Z_{h,\ell}
:=
X_{h,\ell}-\widehat b_hT_{h,\ell}.
\]
The cycle-level renewal-reward standard error is
\begin{equation}
\widehat{\operatorname{SE}}(\widehat b_h)
=
\frac{1}{\sqrt{n_h}\,\overline T_h}
\left[
\frac{1}{n_h-1}
\sum_{\ell=1}^{n_h}
Z_{h,\ell}^{\,2}
\right]^{1/2}.
\label{eq:physical-rate-se}
\end{equation}
The one-hop rate is evaluated analytically as
\[
b_1=rp,
\]
because no swapping is required and every newly generated elementary pair in the matched physical implementation satisfies the fidelity threshold.
Every admissible simulated row contains at least \(51865\) cycles with \(X_{h,\ell}=1\), so no sparse- or zero-event uncertainty construction is used for the reported positive rates.
The zero rows reported below are fixed by the deterministic fidelity ceilings above rather than by zero observed Monte Carlo events.

\paragraph{Hop profiles and derived quantities.}
The elementary fidelity at herald arrival and analytical one-hop rate are
\[
C(0)\simeq0.9979123677,
\qquad
b_1=rp\simeq1061.736707\,\mathrm{s}^{-1}.
\]
Table~\ref{tab:physical-hop-profile} reports \(b_h\) and one cycle-level renewal-reward standard error for each simulated hop.

\begin{table}[t]
\centering
\caption{Fidelity-qualified Bell-pair delivery rates \(b_h\) for isolated sessions at \(a_N=a=1\,\mathrm{km}\), \(1/r=45\,\mu\mathrm{s}\), and \(F_\star=0.95\). Parentheses give one cycle-level renewal-reward standard error. Each simulated hop uses four replicas totaling \(10^6\) shared-history cycles for each physical implementation. The noisy physical implementation zeros at \(h=10,11,12\) and all zeros for \(h\ge13\) follow from the deterministic fidelity ceilings, although the \(h=10,11,12\) noisy physical implementation histories are also evaluated in the common run.}
\label{tab:physical-hop-profile}
\begin{tabular}{c c c c c}
\toprule
\(h\)
& \multicolumn{2}{c}{\shortstack{Noisy physical\\implementation}}
& \multicolumn{2}{c}{\shortstack{One-round DEJMPS\\physical implementation}}\\
\cmidrule(lr){2-3}\cmidrule(lr){4-5}
& \(b_h\,[\mathrm{s}^{-1}]\)
& cycles
& \(b_h\,[\mathrm{s}^{-1}]\)
& cycles\\
\midrule
1
& \(1061.736707\)
& analytic
& \(1061.736707\)
& analytic
\\
2
& \(621.28\;(0.40)\)
& \(1.0\times10^6\)
& \(600.77\;(0.39)\)
& \(1.0\times10^6\)
\\
3
& \(521.34\;(0.29)\)
& \(1.0\times10^6\)
& \(493.96\;(0.28)\)
& \(1.0\times10^6\)
\\
4
& \(465.28\;(0.24)\)
& \(1.0\times10^6\)
& \(435.82\;(0.22)\)
& \(1.0\times10^6\)
\\
5
& \(429.00\;(0.20)\)
& \(1.0\times10^6\)
& \(399.38\;(0.18)\)
& \(1.0\times10^6\)
\\
6
& \(402.30\;(0.18)\)
& \(1.0\times10^6\)
& \(373.60\;(0.16)\)
& \(1.0\times10^6\)
\\
7
& \(375.79\;(0.18)\)
& \(1.0\times10^6\)
& \(353.60\;(0.15)\)
& \(1.0\times10^6\)
\\
8
& \(256.80\;(0.23)\)
& \(1.0\times10^6\)
& \(314.27\;(0.15)\)
& \(1.0\times10^6\)
\\
9
& \(19.038\;(0.081)\)
& \(1.0\times10^6\)
& \(187.67\;(0.15)\)
& \(1.0\times10^6\)
\\
10
& analytic zero
& \(1.0\times10^6\)
& \(93.63\;(0.13)\)
& \(1.0\times10^6\)
\\
11
& analytic zero
& \(1.0\times10^6\)
& \(42.093\;(0.099)\)
& \(1.0\times10^6\)
\\
12
& analytic zero
& \(1.0\times10^6\)
& \(15.694\;(0.065)\)
& \(1.0\times10^6\)
\\
\(\ge13\)
& analytic zero
& ---
& analytic zero
& ---
\\
\bottomrule
\end{tabular}
\end{table}

Using the main-text notation \(\widehat\Lambda\) for the Monte Carlo estimate of \(\Lambda\), summing the unrounded \(h\widehat b_h\) values over the complete admissible support gives
\[
\widehat\Lambda_{\mathrm{no\,pur}}=15144.5\,\mathrm{s}^{-1},
\qquad
\widehat\Lambda_{\mathrm{DEJMPS}}=17992.9\,\mathrm{s}^{-1}.
\]
Independence across hops allows the variances of \(h\widehat b_h\) to be added.
The resulting renewal-reward standard errors are
\[
3.2\,\mathrm{s}^{-1}
\qquad\text{and}\qquad
3.4\,\mathrm{s}^{-1},
\]
respectively.
The nominal asymptotic-normal \(95\%\) intervals, computed from the unrounded estimates and standard errors, are
\[
\Lambda_{\mathrm{no\,pur}}\in[15138.3,\,15150.7]\,\mathrm{s}^{-1},
\qquad
\Lambda_{\mathrm{DEJMPS}}\in[17986.3,\,17999.6]\,\mathrm{s}^{-1}.
\]
These intervals quantify Monte Carlo uncertainty at the fixed physical parameters.

For the matched \(w=1\) resource reference,
\[
r\Gamma_{1,p}
=\frac{rp}{(1-p)^2}
\simeq1170.955925\,\mathrm{s}^{-1}.
\]
Because \(C_{\mathrm{ET}}^{(1)}(N,M;0.95)=C_{\mathrm{ET}}^{(1)}(N,M;1)\) exactly at this benchmark point, the estimated large-\(N\) limits of \(\underline S_{\mathrm{ET}}/(rC_{\mathrm{ET}}^{(1)})\) are
\[
\frac{\widehat\Lambda_{\mathrm{no\,pur}}}{r\Gamma_{1,p}}
\simeq12.9335,
\qquad
\frac{\widehat\Lambda_{\mathrm{DEJMPS}}}{r\Gamma_{1,p}}
\simeq15.3660,
\]
with nominal asymptotic-normal \(95\%\) intervals
\[
[12.9282,12.9388]
\qquad\text{and}\qquad
[15.3603,15.3717],
\]
respectively.
These values estimate the large-\(N\) limits in Eq.~\eqref{eq:app-physical-single-slot-ratio}, rather than exact finite-\(N\) service ratios.

The difference between the two physical hop-profile coefficients is estimated as
\[
\Delta\widehat\Lambda
:=
\widehat\Lambda_{\mathrm{DEJMPS}}-\widehat\Lambda_{\mathrm{no\,pur}}
\simeq2848.4\,\mathrm{s}^{-1},
\]
corresponding to an estimated relative increase of \(18.8\%\).
Using the covariance from the common Monte Carlo histories gives a covariance-corrected standard error of \(2.7\,\mathrm{s}^{-1}\) and the nominal normal \(95\%\) interval
\[
\Lambda_{\mathrm{DEJMPS}}-\Lambda_{\mathrm{no\,pur}}
\in[2843.2,\,2853.6]\,\mathrm{s}^{-1}.
\]
The positive coefficient difference is therefore statistically resolved under the stated physical implementations.
The one-round DEJMPS physical implementation lowers \(b_h\) for \(h=2,\ldots,7\), raises \(b_h\) for \(h=8,9\), and has \(b_h>0\) for \(h=10,11,12\).
Since \(b_1=rp>0\) and both hop profiles have finite support, Eq.~\eqref{eq:app-physical-chain-summable} applies to both physical implementations.

\paragraph{Reproducibility and independent checks.}
The reproducibility package for the terrestrial physical-layer benchmark is archived on Zenodo at \href{https://doi.org/10.5281/zenodo.23031358}{doi:10.5281/zenodo.23031358}, with release version \texttt{1.1.0} and calculation configuration \texttt{20260905.1}.
The archive records the parameters, seed construction, software versions, per-replica sufficient statistics, numerical precision, validation records, and file hashes.
Full reference reruns reproduce the deterministic scientific outputs byte for byte with the pinned numerical-library versions, including across distinct Python environments; environment-specific metadata need not be byte-identical.
These reruns establish deterministic reproducibility of the recorded computation, not physical correctness or statistical convergence outside the tested parameter point.
Independent state-vector and density-matrix calculations check the DEJMPS, dephasing, and swapping primitives.
Closed-form, bisection, and 70-digit evaluations of the waiting budgets agree to within \(2\times10^{-11}\) activation slots, and the waiting-budget criterion reproduces the simulated fidelity qualification of each of the first \(20000\) reference histories per hop.
A separate analytical two-hop calculation gives \(b_2^{\mathrm{no\,pur}}\simeq620.634493\,\mathrm{s}^{-1}\), within \(1.62\) Monte Carlo standard errors of the simulated value.
The same analytical calculation gives an unqualified-output probability of approximately \(8.29\times10^{-22}\), so observing only fidelity-qualified Bell-pair outputs at this hop does not establish probability one analytically.

\subsection{Ideal-retention physical implementation}
\label{app:ideal-retention}

\paragraph{Derivation of \(b_h\).}
The ideal-retention physical implementation keeps the matched link opportunity probability and independence assumptions, assumes ideal preparation, memories, and operations, assumes zero swapping duration, neglects path-wide signaling delays, and retains at most one elementary pair per link location per delivery cycle.
Each link location retains its first successful elementary pair without degradation until every link location on the session path has supplied one elementary pair, after which ideal swapping produces one fidelity-qualified Bell pair.
Here the retained elementary pair at link location \(j\) is its first successful one, so its heralding activation slot \(\tau_j\) is geometric with success probability \(p\) and support \(\{1,2,\ldots\}\), and \(\tau_1,\ldots,\tau_h\) are independent.
The first activation slot at whose end every link location has supplied an elementary pair is
\[
\tau_{\mathrm{ready}}=\max_{1\le j\le h}\tau_j,
\]
whose expectation is exactly
\begin{equation}
\mathbb E[\tau_{\mathrm{ready}}]
=
\sum_{k=0}^{\infty}
\left[
1-
\left(
1-(1-p)^k
\right)^h
\right].
\label{eq:ideal-retention-max-geometric}
\end{equation}
An equivalent exact expression was derived by Bernardes et al.~\cite[Appendix B]{bernardes2011rate} and subsequently treated by Shchukin et al.~\cite[Eq.~(2)]{shchukin2019waiting}.
The summand in Eq.~\eqref{eq:ideal-retention-max-geometric}, extended to a real argument \(x\ge0\), decreases from one to zero and satisfies
\[
\int_0^\infty
\left[1-\left(1-(1-p)^x\right)^h\right]\,dx
=
\frac{\sum_{j=1}^{h}j^{-1}}{-\log(1-p)}.
\]
Comparison of the sum with this integral gives
\[
\frac{\sum_{j=1}^{h}j^{-1}}{-\log(1-p)}
\le
\mathbb E[\tau_{\mathrm{ready}}]
\le
1+\frac{\sum_{j=1}^{h}j^{-1}}{-\log(1-p)}.
\]
Since \(\sum_{j=1}^{h}j^{-1}=\log h+O(1)\), for fixed \(p\in(0,1)\) we obtain
\begin{equation}
\mathbb E[\tau_{\mathrm{ready}}]
=
\frac{\log h}{-\log(1-p)}
+
O(1),
\qquad
h\to\infty.
\label{eq:ideal-retention-waiting}
\end{equation}
The logarithmic growth at fixed \(p\) is also identified by Shchukin et al.~\cite[Sec.~III]{shchukin2019waiting}.
Because one activation slot has reference duration \(1/r\) and one fidelity-qualified Bell pair is delivered per delivery cycle,
\begin{equation}
b_h
=
\frac{r}{\mathbb E[\tau_{\mathrm{ready}}]}
=
\Theta\!\left(\frac{rp}{\log h}\right),
\qquad
(\log h)\,b_h
\longrightarrow
r[-\log(1-p)],
\qquad h\to\infty,
\label{eq:ideal-retention-hop-rate}
\end{equation}
for fixed \(p\in(0,1)\).
Thus
\[
\sum_{h\ge1}h b_h=\infty,
\]
so Eq.~\eqref{eq:app-physical-chain-summable} does not apply.

\paragraph{Multi-user service.}
For the remainder of this subsection, let \(N\to\infty\) in expansion with \(a_N=a\) and fixed \(p\) and \(r\), uniformly over even \(2\le M\le N\).
Apply the construction with weights \(a_Nh_i b_{h_i}\).
A lower bound follows because selecting any one specified demand session is feasible and the construction performs at least as well.
The marginal hop distance \(h_1\) has probability bounded away from zero of lying between \(N/4\) and \(N/2\), and Eq.~\eqref{eq:ideal-retention-hop-rate} therefore gives
\begin{equation}
\mathbb E_{\mathcal D}\!\left[a_Nh_1b_{h_1}\right]
=
\Omega\!\left(
\frac{rpa_NN}{\log N}
\right).
\label{eq:ideal-retention-lower}
\end{equation}
For the matching upper bound, split the sessions selected by the construction into those with \(h_i\le N^{1/3}\) and those with \(h_i>N^{1/3}\).
Since \(\tau_{\mathrm{ready}}\ge\tau_1\), we have \(b_h\le rp\), and Eq.~\eqref{eq:chain-demand-identity} bounds the expected contribution of all short demand sessions, even before enforcing pairwise link-disjointness, by
\begin{align}
\mathbb E_{\mathcal D}\!\left[
\sum_{i:h_i\le N^{1/3}}a_Nh_i b_{h_i}
\right]
&\le
\frac{Ma_Nrp}{N(N-1)}
\sum_{h\le N^{1/3}}(N-h)h \nn\\
&=
O\!\left(rpa_NN^{2/3}\right)
=
o\!\left(
\frac{rpa_NN}{\log N}
\right).
\label{eq:ideal-retention-short}
\end{align}
For the selected long sessions, pairwise link-disjointness implies that their total hop length is at most \(N-1\).
Equation~\eqref{eq:ideal-retention-hop-rate} gives \(b_h=O(rp/\log N)\) uniformly for \(h>N^{1/3}\), and hence their total service is deterministically bounded by
\begin{equation}
\sum_{i:h_i>N^{1/3}}a_Nh_i b_{h_i}
=
O\!\left(
\frac{rpa_NN}{\log N}
\right).
\label{eq:ideal-retention-long}
\end{equation}
Combining Eqs.~\eqref{eq:ideal-retention-lower}--\eqref{eq:ideal-retention-long} yields
\begin{equation}
\underline S_{\mathrm{ET}}
=
\Theta\!\left(
\frac{rpa_NN}{\log N}
\right).
\label{eq:app-ideal-retention-service}
\end{equation}
The absence of \(M\) from this order follows because a single random session has hop distance \(\Theta(N)\) with nonvanishing probability, while pairwise link-disjoint selected sessions have total hop length at most \(N-1\).
At the same \(w=W=1\), \(F_\star=0.95\) benchmark point, Proposition~\ref{prop:fq-single-slot} and the perfect-output single-slot asymptotic give
\begin{equation}
\frac{\underline S_{\mathrm{ET}}}
{rC_{\mathrm{ET}}^{(1)}}
=
\Theta\!\left(
\frac{N^2}{M\log N}
\right).
\label{eq:ideal-retention-single-slot-ratio}
\end{equation}

\paragraph{Relation to asymptotic aggregation.}
The ideal-retention physical implementation removes physical degradation but still retains at most one elementary pair per link location per delivery cycle.
Earlier successful link locations therefore cannot accumulate additional elementary pairs for later delivery cycles while waiting for the last link location to succeed, and Eq.~\eqref{eq:ideal-retention-waiting} produces the logarithmic time penalty.
By contrast, asymptotic aggregation can accumulate elementary pairs supplied by realized links over arbitrarily long resource windows and allocate their average rates through the chain flow problem.
At reference slot rate \(r\), the resource-layer capacity satisfies
\[
rC_{\mathrm{ET}}^{(\infty)}(N,M;0.95)
=
\Theta(rpa_NN)
\]
for \(w=1\) under the same fixed-parameter expansion regime.
The ideal-retention physical implementation therefore achieves service a factor \(\Theta(\log N)\) below its matched resource-layer ceiling \(rC_{\mathrm{ET}}^{(\infty)}(N,M;0.95)\).

\end{document}